\documentclass[11pt,a4paper,UKenglish]{article}
\pdfoutput=1
\usepackage{babel}  
\usepackage{jcappub}
\usepackage{graphicx,cancel}
\usepackage[normalem]{ulem}
\usepackage{txfonts}
\usepackage{relsize}
\usepackage{tikz}
\usepackage{numprint}
\usetikzlibrary{arrows,decorations.pathmorphing,decorations.markings,backgrounds,positioning,fit,calc,shapes,shapes,matrix}
\usepackage{nicefrac}
\usepackage{gensymb}
\usepackage{amsmath}
\usepackage{afterpage}
\usepackage[section]{placeins}
\usepackage{varioref}
\definecolor{webblue}{rgb}{0,0,0.6}
\definecolor{purple}{rgb}{0.5,0,.5}
\definecolor{webgreen}{rgb}{0,0.5,0}
\definecolor{burgundy}{rgb}{0.5, 0.0, 0.13}
\definecolor{evilblue}{rgb}{0.05,0.05,0.35}
\definecolor{darkslategray}{rgb}{0.18, 0.31, 0.31}
\definecolor{darkslategray2}{rgb}{0.21, 0.32, 0.5}
\definecolor{amber}{rgb}{1.0,0.49,0.0}
\definecolor{buddhistmonkrobe}{rgb}{0.966,0.23,0.0}
\definecolor{alizarin}{rgb}{0.82,0.1,0.26}
\definecolor{cadmiumgreen}{rgb}{0.0, 0.42, 0.24}
\definecolor{hanpurple}{rgb}{0.32, 0.09, 0.98}
\definecolor{lightbrown}{rgb}{0.8,0.4,0}
\definecolor{limegreen}{rgb}{0.2, 0.8, 0.2}
\definecolor{limestone}{rgb}{.46,0.46,0.42}
\definecolor{dirtyred}{rgb}{0.31,0.042,0.0625}
\definecolor{silverblue}{rgb}{0.42,0.45,0.48}
\definecolor{sandstone}{rgb}{0.69,0.69,0.59}
\definecolor{armygreen}{rgb}{0.29, 0.32,0.12}
\definecolor{chromeyellow}{rgb}{1.0, 0.65, 0.0}
\definecolor{azure}{rgb}{0.0, 0.5, 1.0}
\definecolor{lilac}{rgb}{0.7,0.58,0.75}
\renewcommand{\textcolor}[1]{} 
\newcommand{\jpr}[1]{\textcolor{buddhistmonkrobe}{#1}}

\newcommand{\done}[1]{{#1}} %

\newcommand{\editorial}[1]{}
\newcommand{\commentinv}[1]{}
\renewcommand{\sout}[1]{} 
\newcommand{\soutinv}[1]{}
\newcommand\dd{\mathrm{d}}
\newcommand\ii{\mathrm{i}}
\newcommand\ee{\mathrm{e}}

\newcommand\deriv[2]{ \displaystyle\frac{\partial #1}{\partial #2} } 
\newcommand{\RM}{\mathrm{RM}}       
\newcommand{\rad}{\mathrm{rad}}		
\newcommand{\cm}{{\rm cm}}    
\newcommand{\km}{{\rm km}}    
\newcommand{\m}{{\rm m}}      
\newcommand{\pc}{{\rm pc}}    
\newcommand{\kpc}{{\rm kpc}}  
\newcommand{\Mpc}{{\rm Mpc}}  
\newcommand{\g}{{\rm g}}      
\newcommand{\Msun}{\mathrm{M}_{\odot}} 
\newcommand{\s}{{\rm s}}      
\newcommand{\Myr}{{\rm Myr}}  
\newcommand{\Gyr}{{\rm Gyr}}  
\newcommand{\MHz}{{\rm MHz}}  
\newcommand{\GHz}{{\rm GHz}}  
\newcommand{\kms}{\km/\s}    
\newcommand{\muG}{{\rm \mu G}} 
\newcommand{\nG}{{\rm nG}}    
\newcommand{\EV}{{\rm EV}}    

\newcommand{\eV}{{\rm eV}}    
\newcommand{\GeV}{{\rm GeV}}  
\newcommand{\TeV}{{\rm TeV}}  
\newcommand{\PeV}{{\rm PeV}}  
\newcommand{\EeV}{{\rm EeV}}  

\newcommand{\ie}{i.e.,} %

\labelformat{chapter}{chapter~#1}
\labelformat{section}{section~#1}
\labelformat{appendix}{appendix~#1}
\labelformat{subsection}{section~#1}
\labelformat{subsubsection}{section~#1}
\labelformat{figure}{figure~#1}
\labelformat{subfigure}{figure~\thefigure #1}
\labelformat{table}{table~#1}
\labelformat{equation}{eq.~(#1)}

\newcommand{\textsw}[1]{\textsl{#1}}
\newcommand{\hammurabi}{\textsw{hammurabi}}
\newcommand{\hammurabiX}{\textsw{hammurabi\,X}}
\newcommand{\healpix}{\textsw{HEALPix}}

\newcommand{\pymc}{\textsw{PyMC}}
\newcommand{\pymcthree}{\textsw{PyMC\,3}}
\newcommand{\pymultinest}{\textsw{PyMultiNest}}
\newcommand{\stan}{\textsw{STAN}}
\newcommand{\nifty}{\textsw{NIFTy}}
\newcommand{\niftythree}{\textsw{NIFTy\,3}}
\newcommand{\dtwoo}{\textsw{D2O}}
\newcommand{\python}{\textsw{Python}}

\newcommand{\crpropathree}{\textsw{CRPropa\,3}}
\newcommand{\imagine}{\textit{IMAGINE}}
\newcommand{\imagineC}{IMAGINE Consortium}
\newcommand{\imagineSW}{\textsw{IMAGINE}}
\newcommand{\planck}{\textit{Planck}}
\newcommand{\planckC}{Planck Collaboration}

\hypersetup{
pdfauthor={IMAGINE Consortium},%
pdftitle={IMAGINE: A comprehensive view of the ISM, GMFs and CRs}%
}

\renewcommand{\citet}{\cite}
\renewcommand{\citealt}{\cite}

\newcounter{legalnote}
\begin{document}
   \title{IMAGINE: A comprehensive view of the interstellar medium, Galactic magnetic fields and cosmic rays}
   \collaboration{IMAGINE Consortium}
   \author[1,\dag]{Fran\c{c}ois~Boulanger,\note[\dag]{corresponding author}}
   \author[2,\dag]{Torsten~En{\ss}lin,}
   \author[3]{Andrew~Fletcher,}
   \author[4]{Philipp~Girichidis,}
   \author[5]{Stefan~Hackstein,}
   \author[6,\dag]{Marijke~Haverkorn,}
   \author[6,7,\dag]{J\"org~R.~H\"orandel,}
   \author[8,9,\dag]{Tess~Jaffe,}
   \author[10,\dag]{Jens~Jasche,}
   \author[11]{Michael~Kachelrie{\ss},}
   \author[12]{Kumiko~Kotera,}
   \author[4]{Christoph~Pfrommer,}
   \author[6,\dag]{J\"org~P.~Rachen,}
   \author[3]{Luiz~F.~S.~Rodrigues,}
   \author[13,14]{Beatriz~Ruiz-Granados,}
   \author[3]{Amit~Seta,}
   \author[3,\dag]{Anvar~Shukurov,}
   \author[15]{G{\"u}nter~Sigl,}
   \author[2]{Theo~Steininger,}
   \author[16]{Valentina~Vacca,}
   \author[6,17]{Ellert~van~der~Velden,}
   \author[6]{Arjen~van~Vliet,}
   \author[18,19]{and Jiaxin~Wang}

\affiliation[1]{LERMA/LRA, Observatoire de Paris, PSL Research University, CNRS, Sorbonne Universit\'e,\hspace{\fill}\linebreak \mbox{UPMC Universit\'e Paris 06}, Ecole
Normale Sup\'erieure, 75005 Paris, France}
\affiliation[2]{Max Planck Institute for Astrophysics, Karl-Schwarschildstr. 1, 85741 Garching, Germany}
\affiliation[3]{School of Mathematics, Statistics and Physics, Newcastle University, Newcastle upon Tyne,\\ \mbox{NE1 7RU, UK}}
\affiliation[4]{Leibniz-Institut f\"{u}r Astrophysik Potsdam,  An der Sternwarte 16, 14482 Potsdam, Germany}
\affiliation[5]{Hamburger Sternwarte, Gojenbergsweg 112, 21029 Hamburg, Germany}
\affiliation[6]{Department of Astrophysics/IMAPP, Radboud University,  \jpr{PO Box 9010, 6500 GL Nijmegen, \sout{Heyendaalseweg 135, \mbox{6525 AJ Nijmegen}}}, \mbox{The Netherlands}}
\affiliation[7]{Nikhef, Science Park \jpr{105, 1098 XG} Amsterdam, The Netherlands}
\affiliation[8]{CRESST II, NASA Goddard Space Flight Center, Greenbelt, MD, 20771, USA}
\affiliation[9]{Department of Astronomy, University of Maryland, College Park, MD, 20742, USA}
\affiliation[10]{The Oskar Klein Centre, Department of Physics, Stockholm University, \mbox{AlbaNova University Centre}, SE 106 91 Stockholm, Sweden}
\affiliation[11]{Institutt for fysikk, NTNU, Trondheim, Norway}
\affiliation[12]{Sorbonne Universit\'{e}, UPMC Univ.\ Paris 6 et CNRS, UMR 7095, Institut d'Astrophysique de Paris, 98 bis bd Arago, 75014 Paris, France}
\affiliation[13]{Instituto de Astrof\'{i}sica de Canarias. V\'{i}a L\'{a}ctea, s/n.\ E-38205, La Laguna, Tenerife, Spain} 
\affiliation[14]{Departamento de Astrof\'{i}sica.\ Universidad de La Laguna, E-38206, La Laguna, Tenerife, Spain}
\affiliation[15]{Universit\"at Hamburg, {II}.\ Institute for Theoretical Physics, Luruper Chaussee 149, 22761 Hamburg, Germany}
\affiliation[16]{INAF, Osservatorio Astronomico di Cagliari, Via della Scienza 5, I-09047 Selargius, Italy}
\affiliation[17]{Centre for Astrophysics and Supercomputing, Swinburne University of Technology, PO Box 218, Hawthorn, VIC 3122, Australia}
\affiliation[18]{Scuola Internazionale Superiore di Studi Avanzati, Via Bonomea 265, 34136 Trieste, Italy}
\affiliation[19]{Istituto Nazionale di Fisica Nucleare, Sezione di Trieste, Via Bonomea 265, 34136 Trieste, Italy}

\emailAdd{francois.boulanger@ias.u-psud.fr}
\emailAdd{ensslin@mpa-garching.mpg.de} 
\emailAdd{m.haverkorn@astro.ru.nl} 
\emailAdd{j.horandel@astro.ru.nl}
\emailAdd{tess.jaffe@nasa.gov}
\emailAdd{jens.jasche@fysik.su.se}
\emailAdd{j.rachen@astro.ru.nl} 
\emailAdd{anvar.shukurov@newcastle.ac.uk}


\abstract{
In this white paper we introduce the \imagineC\ and its scientific 
background, goals and structure. The purpose 
of the consortium is to 
coordinate and facilitate the efforts of a diverse group of researchers in the 
broad areas of the interstellar medium, Galactic magnetic fields and cosmic rays, and our 
overarching goal is to develop more comprehensive insights into the structures and roles of 
interstellar magnetic fields and their interactions with cosmic rays within the 
context of Galactic astrophysics.
The ongoing rapid development of observational 
and numerical facilities and techniques has resulted in a widely felt need to 
advance this subject to a qualitatively higher level of self-consistency, depth 
and rigour. This can only be achieved by the coordinated efforts of experts in 
diverse areas of astrophysics involved in observational, theoretical and numerical 
work. We present our view of the present status of this research area, identify 
its key unsolved problems and suggest a strategy that will underpin our work. The 
backbone of the consortium is the \textit{Interstellar MAGnetic field INference 
Engine}, a publicly available Bayesian platform that employs robust statistical 
methods to explore the multi-dimensional likelihood space using any number of modular inputs.  
This tool will be used by the \imagineC\ to develop an interpretation and modelling 
framework that provides the method, power and flexibility to interfuse information 
from a variety of observational, theoretical and numerical lines of evidence into 
a self-consistent and comprehensive picture of the thermal and non-thermal 
interstellar media. An important innovation is that a consistent understanding of the phenomena that are directly or indirectly influenced by the Galactic magnetic field, such as the deflection of ultra-high energy cosmic rays or extragalactic backgrounds, is made an integral part of the modelling. The \imagineC, which is informal by nature and open to new 
participants, hereby \done{presents} a methodological \done{framework} for the modelling and understanding of Galactic magnetic fields that is available to all communities whose research relies on a state of the art solution to this problem.}

   \arxivnumber{1805.02496}
   \dedicated{A tribute to John Lennon}
   \maketitle
    \flushbottom

\section{Introduction}
\label{IES}

\subsection{IMAGINE aims} 
In many areas of astrophysics, the magnetic fields are the least understood component and yet one of the most important. From the cosmological question of structure formation to the shapes of dusty filaments in the interstellar medium (ISM), magnetic fields on all scales play a critical role that has often been neglected. In the Milky Way, the Galactic magnetic fields (GMFs) affect all phases of the ISM from the propagation of relativistic cosmic rays to the collapse of cold dust clouds. Beyond our galaxy, the detection of primordial magnetic fields, the role of magnetic feedback in galaxy formation, and the structure of extragalactic large-scale magnetic fields are outstanding questions in cosmology and large-scale structure formation. But magnetic fields are not only important in their dynamical role; for many \done{research areas} they also turn out to be a nuisance. The polarised Galactic foregrounds are the main challenge for the detection of primordial gravitational waves with the cosmic microwave background (CMB), and for ultra-high energy cosmic rays (UHECRs), Galactic and extragalactic magnetic fields blur our view of their sources.

In nearly every context, the properties of the magnetic fields and particles as well as the properties and dynamics of the emergent structures are entangled and interdependent. Studies of any one piece of the puzzle alone are bound to be limited in scope and likely to be biased in their results. Until recently, that was the best we could do to make the problem tractable. But we now have the observational, computational, and mathematical tools to start to bring all these threads together. The different contexts and different effects provide complementary information that we can now combine. 

The \imagineC\ was conceived to join the experts in these disparate topics into a single team for a comprehensive study of the Galactic magnetic field. We will use all available data, from traditional observables such as synchrotron emission and Faraday rotation measures, to newer tracers such as polarised thermal dust emission and UHECR deflections. To model the components of the magnetised ISM, we will use not only heuristic models of the Galaxy motivated by observations, e.g., with spiral arm segments and field reversals, but also theoretically motivated parametric models and finally non-parametric models constrained by fundamental magnetohydrodynamics (MHD). We will further include constraints from dynamo theory and knowledge gained from observations of nearby galaxies.

\begin{figure*}[tb]
\includegraphics[width=\textwidth]{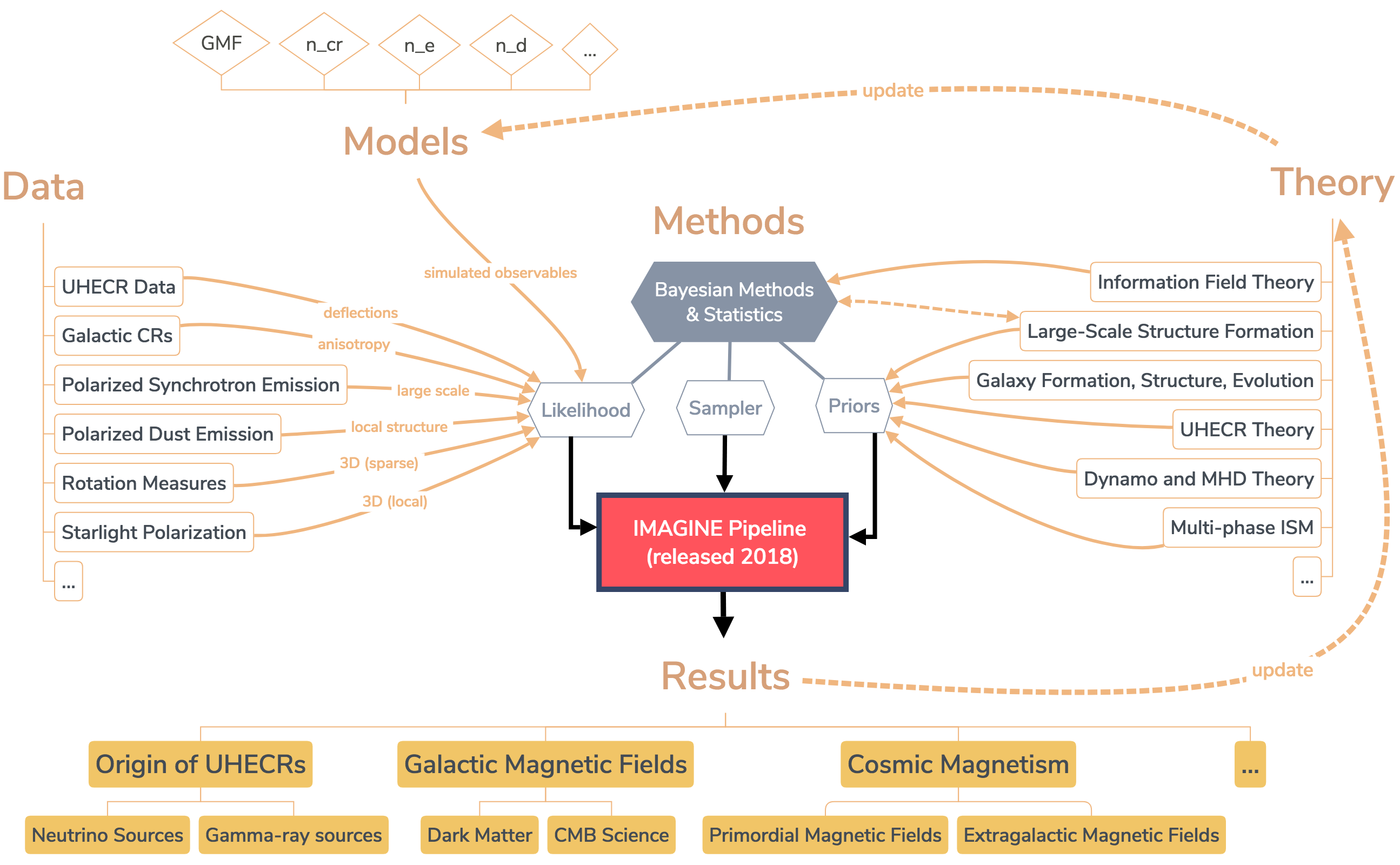}
\caption{\label{fig:imagine_mindmap} Structure of the \imagine\ project.  See text for further explanation.}
\end{figure*}

The \imagine\ project is based on Bayesian methods, which are particularly important for solving interdependent problems, e.g., the origin \textit{and} deflection of UHECRs. They also provide a robust quantification of what we have learned with each additional piece of the analysis puzzle. Our goal is to provide a standard framework for updating our knowledge of the structure of the GMF with new data from next generation observational facilities, new analysis techniques, and advances in theoretical understanding.  The core of the project is the recently released \imagineSW\ pipeline \citep{steininger2018}, a publicly available and modular software infrastructure. An overview of the \imagine\ project is depicted in \ref{fig:imagine_mindmap} and its science goals briefly outlined in the following paragraphs. The current paper is thought of as a white paper for the \imagine\ project and aims to give a summary of the various research fields involved and the impact \imagine\ will have.

\subsection{IMAGINE science}

Members of the \imagineC\ are actively researching in different areas of Galactic science that are important for understanding the GMF.  These include but are not limited to the following:
\begin{itemize}
\item The multi-phase ISM (\ref{ss:MPISM}):  magnetic fields are a crucial component of the ISM ecosystem and have a complex interdependence on the thermal gas and dust, the relativistic particles, the star formation, supernovae, etc.  
\item Galactic cosmic rays (\ref{ss:gCR}): cosmic rays diffuse through the Galaxy, couple to the magnetic fields, and affect the dynamics of the ISM and of the entire Galaxy. 
\item The Galactic magnetic field (\ref{ss:OBSGMF}): a variety of complementary observational tracers tell us about the morphology of the large-scale coherent magnetic field, the statistical and morphological properties of its small-scale turbulent structures, and the interplay between them. 
\item Dynamo and MHD theory (\ref{ss:DTGMF}): MHD theory constrains the morphology of the GMF on both large and small scales and is crucial for understanding the turbulence in the ISM. 
\end{itemize}
All these research areas are described in more detail throughout the paper in the sections referenced above. Beyond this, there are also various extragalactic and cosmological topics that contribute to, or profit from, our understanding of the GMF, such as:
\begin{itemize}
\item Large-scale structure formation (\ref{ss:SF}): the distribution of matter in our local Universe and its formation history affect most extragalactic tracers important for \imagine, such as UHECR sources or extragalactic magnetic fields.  
\item Galaxy formation and evolution (\ref{sec:gal_form}):  feedback processes are important in galaxy formation and evolution, and the role of magnetic fields and cosmic rays is central but not yet well understood. 
\item Ultra-high energy cosmic rays (\ref{ss:UHECR}):  an improved understanding of the GMF will help us trace UHECRs back to their sources, and likewise, UHECR deflections help us trace the GMF. 
\end{itemize}
Finally, there are various Galactic and extragalactic backgrounds that are of interest for fundamental physics, which will greatly profit from an improved understanding of large-scale structure, the GMF, MHD turbulence, etc. These include cosmological topics from primordial magnetic fields and gravitational waves to the epoch of reionisation (\ref{ss:exbkg}),  as well as the potential signal of dark matter annihilation from our own galaxy (\ref{ss:DM}).

\subsection{IMAGINE methods}
To join all these different research fields in the common goal of modelling the GMF,
we have developed the \textit{Interstellar MAGnetic field INference Engine} (also referred to as the \imagineSW\ pipeline), a framework with the power and flexibility to include information from all of these different fields. The \imagineSW\ pipeline is a Bayesian platform that takes advantage of robust statistical methods to explore the multi-dimensional likelihood space using any number of modular inputs (\ref{s:AM}). These inputs include: 
\begin{itemize}
\item {\it data}, \ie\ all possible tracers of magnetic fields, from Faraday rotation measures (RMs) through submillimeter dust polarisation to UHECR deflections (\ref{ss:OI}); 
\item {\it Galaxy simulations}, \ie\ formulations of models for the magnetic fields (parametric, \ref{ss:Para}, or non-parametric, \ref{ss:non-para-models}) and other Galaxy components, and the simulated observables generated from them;  
\item independent {\it likelihood} evaluations (\ref{sec:Bayes:likelihood}) to compare the mock data to the observed data taking full account of any uncertainties; 
\item the mathematical expression of theoretical constraints or knowledge from observations of external galaxies, etc., in the form of Bayesian {\it priors} (\ref{sec:Bayes:prior}); 
\item efficient {\it samplers} for exploring the multi-dimensional likelihood space (\ref{sec:Bayes:num_bayes}); 
\item the computation of the Bayesian {\it evidence} to quantify what we have learned from different models (\ref{sec:Bayes:evidence}).
\end{itemize}
The infrastructure is summarised in \ref{s:SD} and described in detail in \citet{steininger2018} along with its first results. In that paper, we demonstrate the pipeline on both simulated and real data using a simple GMF model and a subset of the available observables. That first application demonstrates not only the technical success of the pipeline but also the complexity of the task. Our preliminary results underline the need for a comprehensive Bayesian approach with physically motivated priors, \ie\ precisely the challenge that we describe here. 

\subsection{Structure of the paper}
This white paper describes the \imagine\ project in detail. In \ref{s:BGMW}, we give an overview of the various topics in Galactic astrophysics that will contribute to and/or benefit from \imagine, while in \ref{s:BGEG} we discuss extragalactic astrophysics and the cosmological connections. In \ref{s:AM}, we discuss the information theoretical framework of \imagine\ and in particular how we harness the power of Bayesian statistics to tackle this challenge. In \ref{s:SD}, we briefly describe the software package we have developed and how it can be used to add inputs (whether these are data, models, or priors) from all of these different fields of astrophysics as well as others we may not yet have foreseen. 

\begin{table}[t]

\caption{\label{acrotab} Acronyms introduced and used in the text, software packages and experiments excluded.}

 \centering
 \begin{tabular}{r@{\qquad}l@{\qquad\qquad}r@{\qquad}l}
 \hline
 CMB & cosmic microwave background & IFT & information field theory\\
 DM & dark matter & ISM & interstellar medium\\
 EoR & epoch of reionisation & MCMC & Markov chain Monte Carlo\\
 GCR & Galactic cosmic ray & MHD & magnetohydrodynamics\\
 GMF & Galactic magnetic field & RM & rotation measure\\
 EGMF & extragalactic magnetic field & UHECR & ultra-high energy cosmic ray \\ \hline
\end{tabular}
\end{table}


\section{Galactic science}
\label{s:BGMW}

The motivation for the \imagine\ project comes from several directions simultaneously, and the most immediate is the need to understand our own galaxy. Here, we review the topics in Galactic astrophysics that will be advanced by \imagine.

\subsection{The multi-phase interstellar medium}
\label{ss:MPISM}

The ISM is conventionally separated into thermal and non-thermal parts, the
latter represented by magnetic fields and cosmic rays 
and too often neglected. With the development of
modern observational and numerical techniques, this division is becoming more and
more inappropriate. The ISM of a spiral galaxy is a complex system whose constituent
parts may be explored in isolation only as a preliminary investigation, often with little
confidence in qualitative or quantitative fidelity.

Magnetic fields and cosmic rays contribute significantly to the structure and
evolution of the ISM and the host galaxy. They affect the accretion of gas by dark
matter (DM) haloes \citep{RdSO10} as well as the outflows and inflows in
galaxies that have already formed \citep{B2009}. Magnetic fields and cosmic rays
can significantly affect galactic outflows (fountains and winds)
through their effect on the multi-phase structure of
the interstellar gas, as they confine hot gas bubbles produced by
supernovae \citep{FMLZ91,HT06}. According to numerical
simulations, a magnetised multi-phase ISM can be more homogeneous than
a purely thermal one \citep{EGSFB17}. Non-thermal effects modify
the self-regulation of star formation in spiral galaxies and its effects on the intergalactic
gas (\citealt{SSS2010,WPP17}, and references therein).
The magnetic contribution to the overall structure of a galactic
gaseous disc is at least as important as that from other sources of interstellar
pressure (\ie\ thermal, turbulent, and cosmic ray pressure terms), as all of
them are of comparable magnitude \citep{P1979,C2005}. Half of the total
interstellar pressure is thus due to non-thermal contributions, and
magnetic fields therefore directly affect the mean gas density. In turn,
this significantly affects the star formation rate. It is therefore
surprising that the role of magnetic fields and cosmic rays in galaxy
evolution has avoided attention for so long \citep{BBT15,RSFB15}. Magnetic
fields also regulate star formation locally
by controlling the collapse and fragmentation of molecular clouds
\citep{ML2009,C2012}. Magnetic fields
contribute to interstellar gas dynamics not only directly but also by
confining cosmic rays \citep{G90,S2002,Shalchi09}. The latter are effectively
weightless and so are capable of driving galactic outflows
\citep{BK00,EEZBMRG08,UPSNES12,BAKG13,Girichidis2016,pakmoretal2016,Simpson2016},
thus providing negative feedback on star formation in galactic discs
\citep{VCB-H05,PPJ12}.

The crucial role that magnetism plays in the ecosystem of a galaxy
has been  known for years, but only recently have advances in detection methods,
technology, and computer power made major leaps forward and allowed comprehensive studies of
galactic magnetic fields. 
As a result, it has become clear that
simple models of large-scale galactic magnetic fields following galactic
spiral arms are utterly inadequate. Recent data from large multi-wavelength
radio-polarimetric surveys have allowed refinement of these models including,
e.g., anisotropic turbulence and/or vertical field components. However, these
models are still data-starved, and inclusion of magnetic field information
through other sources than radio polarimetry is therefore needed.

\subsection{Galactic cosmic rays}
\label{ss:gCR}

Cosmic rays are ionised atomic nuclei, electrons, and positrons, and their elemental abundance resembles roughly the average abundance in the Solar System. They have an energy density of about $1\,\eV/\cm^3$, comparable to that of the Galactic magnetic field. They are
believed to be accelerated mostly by first-order Fermi processes at strong shocks in sources such as supernova remnants, and they then propagate through the Galaxy.

The propagation of Galactic cosmic rays (GCRs) is a combination of advection with the plasma as well as streaming and diffusion. In the ideal MHD approximation, magnetic fields are flux-frozen into the plasma and thus advected with the flow. Cosmic rays are bound to gyrate along individual field lines and are advected alongside the moving plasma. As they propagate, they resonantly excite Alfv\'en waves, which scatter the cosmic rays. As a result, the GCR distribution (partially) isotropizes in the reference frame of Alfv\'en waves, \ie\ low-energy GCRs stream down their gradient \citep{Zweibel:2013}. MHD turbulence maintained at larger scales by other sources can also scatter cosmic rays, redistributing their pitch angles, but leaving their energy unchanged. This can be described as anisotropic diffusion.

From our local perspective in the Milky Way, the GCR propagation is dominated by streaming and diffusion, since the Sun is almost perfectly co-moving with the orbiting ISM around the Galactic centre. Note that this is not the case if cosmic rays travel large distances from their sources to us due to differential rotation of the ISM, which makes advection more important. As mentioned above, GCR protons with energies ${\lesssim} 100\,\GeV$ are dynamically coupled to the ISM via self-generation of and scattering at Alfv\'en waves. Hence, they cannot be treated as ``test particles'' in a static background, as their back-reaction is important for the dynamics of the ISM and the turbulent magnetic field at scales comparable to the Larmor (gyration)
radius. At significantly larger scales, the MHD turbulence is
likely to be affected more weakly and the test particle approximation remains a useful tool in studies of cosmic ray propagation. Inferences of the cosmic ray pressure distribution via frameworks such as \imagineSW\ thus hold the promise to quantify the dynamical impact of cosmic rays on the ISM and, by extension, on important physical processes that are relevant for galaxy formation (see \ref{sec:gal_form}).

\afterpage{
\addtocounter{footnote}{-1}
\begin{figure}[t]
\centering
\includegraphics[width=0.98\columnwidth]{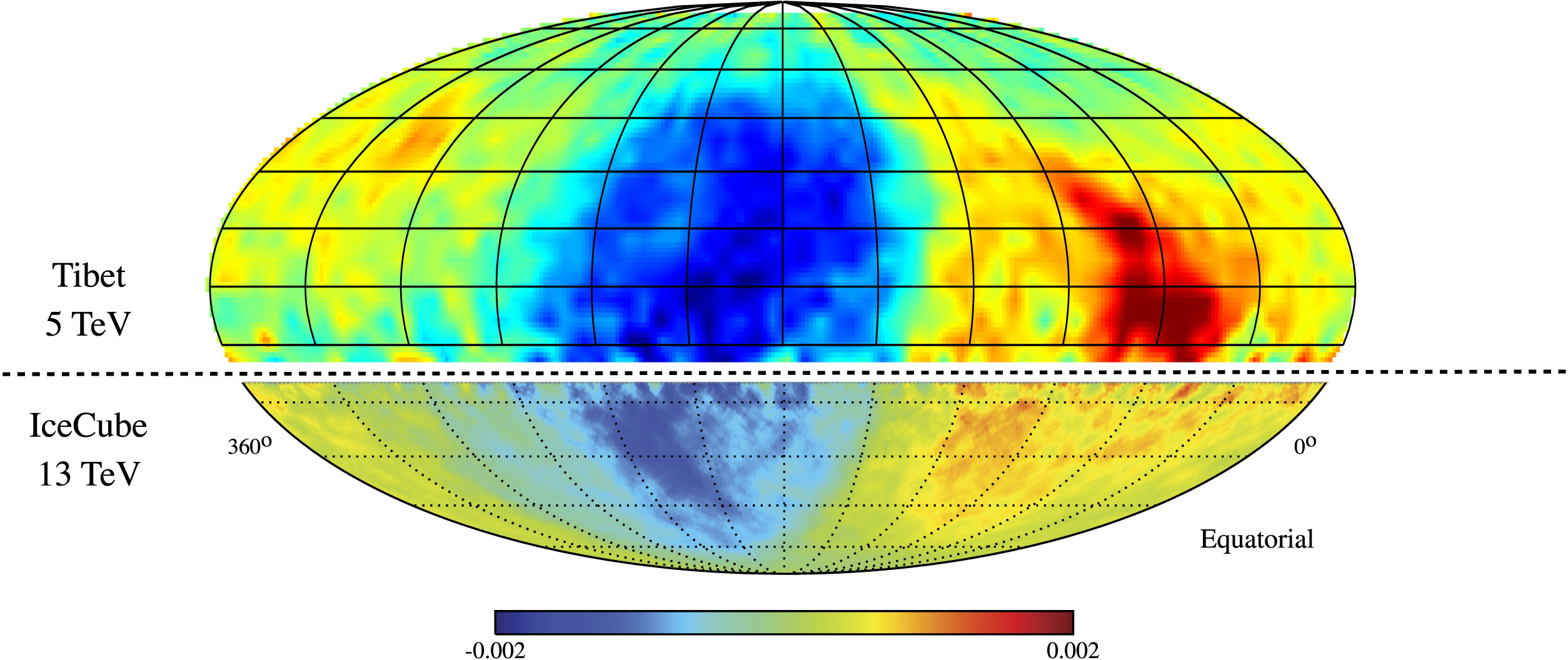}
\caption[Combined cosmic ray anisotropy of the Tibet-AS and IceCube experiments in the equatorial coordinate system.]{\done{Combined cosmic ray anisotropy of the Tibet-AS and IceCube experiments in the equatorial coordinate system. \jpr{\sout{Image credit and detailed information: M.~Ahlers and P.~Mertsch \citet{ahlers:2017}.} See \cite{ahlers:2017} for detailed information.\footnotemark
}}}
\label{fig:CR_anis}
\end{figure}
\footnotetext{\jpr{Reprinted from Progress in Particle and Nuclear Physics, Vol. 94, M.~Ahlers and P.~Mertsch, \textit{Origin of small-scale anisotropies in Galactic cosmic rays}, figure~2, pg.~187, \textcopyright\ (2017), with permission from Elsevier}}
}

At GCR proton energies ${\gtrsim} 100\,\GeV$, the decreasing cosmic ray energy density implies a significantly smaller growth rate of the resonantly excited Alfv\'en waves. Turbulent and non-linear Landau damping leads to a low amplitude of magnetic fluctuations that are not sufficient to self-confine cosmic rays to the Alfv\'en frame, and their propagation becomes mostly diffusive, until the diffusion picture breaks down for cosmic rays with energies ${\gtrsim} 10^{14}\,\eV$. These changes in the mode of propagation manifest themselves in the form of breaks and deviations from the power-law spectrum in rigidity of GCR nuclei. Most importantly in the context of \imagine, the diffusion tensor is connected to the local orientation of the GMF. In an alternative approach, one numerically calculates the trajectories of individual GCR particles, solving the Lorentz equation in the turbulent and regular GMF. It has been demonstrated \citep{escape} that this method allows one to derive global constraints on the properties of the GMF. In particular, the deduced density $n(\vec{x},E)$ of GCR electrons is an important input in the determination of the GMF via synchrotron radiation, while the GMF in turn determines the propagation of cosmic rays. These simple examples illustrate that the properties of the GMF and of GCRs are entangled: deducing these properties therefore requires a joint analysis with a comprehensive approach like \imagineSW.

During propagation, GCRs inelastically interact with nuclei of the ISM, producing (radioactive) nuclei, (anti-)protons, electrons, positrons, and neutrinos as secondary particles in hadronic interactions as well as gamma-ray emission from decaying pions. The secondary electrons/positrons produce secondary radio synchrotron and inverse Compton gamma-ray emission. We can learn about the GCR sources and the diffusion process of cosmic rays in the Milky Way by comparing the modelled primary and the calculated secondary fluxes to observations \citep{SMP}. Observational data  from, e.g., Fermi-LAT, CTA and IceCube, will extend this information to yet higher cosmic ray energies.

These non-thermal radiative GCR and GMF tracers throughout the Galaxy are complemented by cosmic ray measurements at the Earth: the ratio of primary (accelerated at the sources) to secondary (produced through spallation during the propagation processes) cosmic ray nuclei, like the boron-to-carbon ratio, is a measure for the matter traversed by cosmic rays (see e.g., \citealt{obermeier:2012}).
The resulting column density amounts to about $10\,\g/\cm^2$ at $\GeV$ energies, decreasing as a function of energy ${\propto}\, E^{-1/3}$.
At energies around $E=3\cdot 10^{15}\,\eV$, the all-particle energy spectrum of cosmic rays exhibits a change in the spectral index. Measurements indicate that the individual elements in the cosmic ray chemical composition exhibit a fall-off roughly proportional to the rigidity
\citep{antoni:2005,hoerandel:2008}.
The rigidity-dependent fall-off of individual elements is most likely due to a combination of the maximum energy attained in the accelerators and leakage from the Galaxy during the propagation processes
\citep{hoerandel:2004}.
Using gamma-ray astronomy, the fall-off of the energy spectrum can be observed at the sources directly (e.g., \citealt{aharonian:2006}). This provides an opportunity to infer the spectral behaviour of the (hadronic) cosmic rays at the sources. The differences in the fall-offs observed for cosmic rays at the Earth is due to the propagation processes (leakage from the Galaxy), which are in turn connected to the GMF. Recently, several experiments have detected small-scale anisotropies in the arrival directions of $\TeV$ cosmic rays at the level of $10^{-3}$ (\cite{2010ApJ...711..119A, 2016ApJ...826..220A}, 
see \ref{fig:CR_anis}). Explaining these anisotropies requires a detailed understanding of the GMF structure at a very wide range of scales.

\subsection{The Galactic magnetic field}
\label{ss:OBSGMF}
In this section, we will briefly review the main observational tracers of the GMF and the knowledge gained from them. For more extensive reviews, see e.g., \citet{haverkorn2015,kleinfletcher2015,Beck16}.

\subsubsection{Observational tracers}
\label{sss:oot}

There is a rich \done{set} of observational tracers that depend on magnetic fields. However, each of them probes only certain features of the
magnetic field, e.g., only the strength or the direction/orientation,
or only the component parallel or perpendicular to the line-of-sight. In addition, these tracers are sensitive to magnetic fields in
a specific medium, such as cold clouds, diffuse ionized gas, or the
non-thermal synchrotron-emitting plasma. This means that there is no
ideal tracer of GMFs and that {\it all} relevant
tracers should be included to obtain a complete picture. Below, we
briefly discuss the main known observational tracers of the large-scale and
small-scale GMF components.\footnote{We discuss only those tracers for which we have enough data to
probe the diffuse interstellar medium. Tracers
of dense, cold gas and specific (star forming or circumstellar)
environments, such as Zeeman splitting or masers, will not be
discussed here.} In \ref{ss:OI}, we will discuss which specific data sets of these observational tracers will be used as input for \imagineSW.

\afterpage{
\addtocounter{footnote}{-1}
\begin{figure}[t]
\centering
\includegraphics[width=0.85\columnwidth]{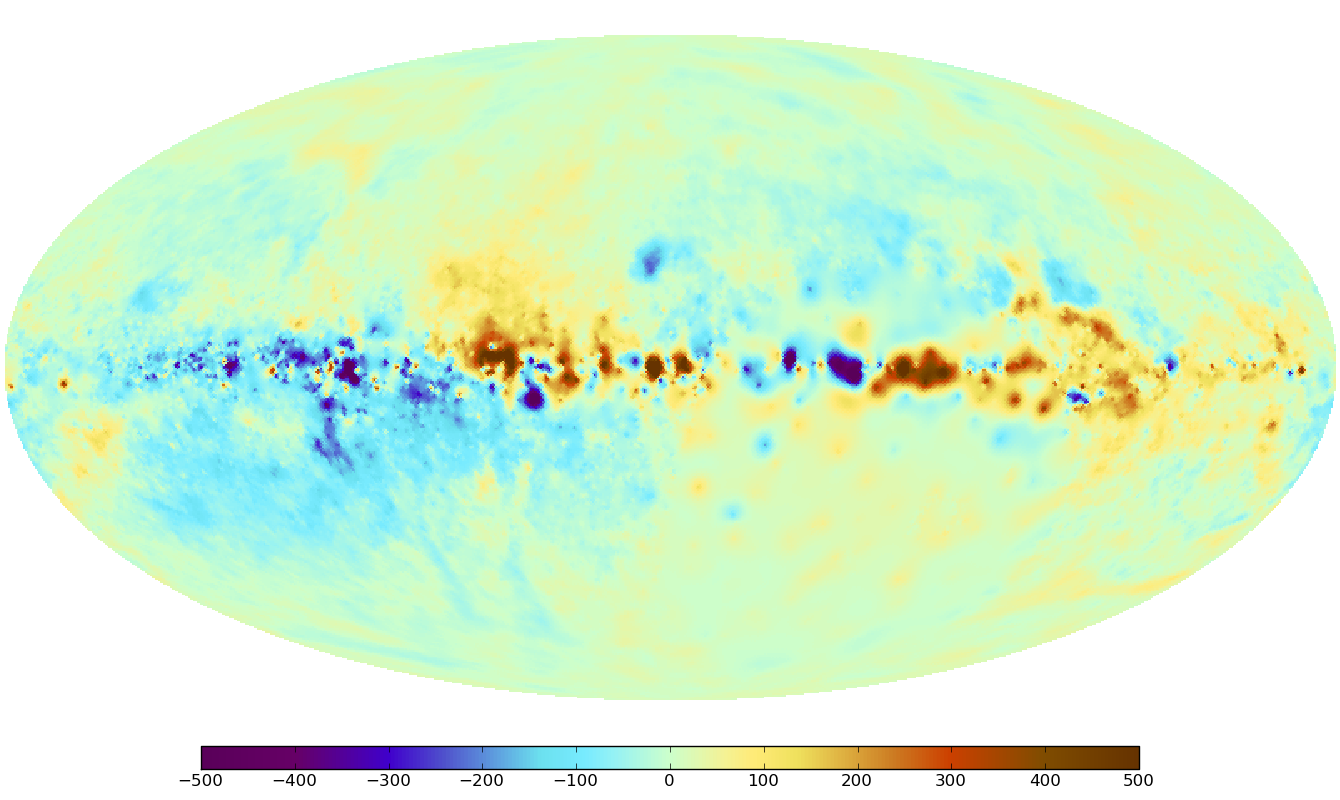}
\caption[All-sky map of Faraday rotation revealing the line-of-sight component of the Galactic magnetic field.]{All-sky map of Faraday rotation by \jpr{Oppermann et al.} \citet{2012A&A...542A..93O} revealing the line-of-sight component of the Galactic magnetic field.\footnotemark}
\label{fig:Oppermann}
\end{figure}
\footnotetext{\jpr{Excerpt from figure~3 in N.~Oppermann et al., A\&A, Vol.~542, A93 (2012), reproduced with permission \textcopyright\ ESO}}
}

\paragraph{Faraday rotation}
\texttt{ }\\
Faraday rotation is the birefringence of left-circularly and right-circularly polarised radiation travelling through a magnetised
plasma. As a result, the polarisation angle of linearly
polarised radiation $\chi$ rotates as a function of wavelength $\lambda$ as
\begin{align*}
\chi &= \chi_0 + \RM\, \lambda^2,
\end{align*}
where $\chi_0$ is the intrinsic polarisation angle with which the
radiation is emitted, and $\RM$ is the rotation measure defined as
\begin{align}
\label{e:rm}
\left(
\frac{\RM}{\rad\,\m^2}
\right) 
&= 0.812
\int_\text{source}^\text{observer}\left(
\frac{n_e(s)}{\cm^{-3}}\right)
\left(\frac{\vec{B}(s)\cdot\dd\vec{s}}{\muG\cdot\pc}\right) ,
\end{align}
where $n_e(s)$ and $\vec{B}(s)$ are the thermal electron density and magnetic vector field in the intervening medium, respectively, at distance $s$ away from the observer.
\done{Extragalactic radio sources therefore allow the construction of images of the total projected GMF (see \ref{fig:Oppermann} and \citealt{2012A&A...542A..93O}), whereas RMs from pulsars at different distances provide tomographic information.}

However, if the synchrotron emitting and Faraday rotating media are
mixed, $\RM$ is no longer given by \ref{e:rm}. Instead we derive a
Faraday spectrum, which is defined as the polarised intensity as a function of
{\it Faraday depth}, $\phi$. Faraday depth is obtained by \ref{e:rm}, with the distinction that now the integration
boundaries vary as a function of distance along the line-of-sight \citep{burn1966,brentjensdebruyn2005}.

Estimates of magnetic field strength from Faraday rotation measure rely on the
assumption that $B_\parallel$ and $n_e$ are uncorrelated, so 
$B_\parallel\propto\RM/(\langle n_e\rangle L)$, where $L$ is the line-of-sight extent of the magneto-ionic medium and $\langle n_e\rangle$ is the average electron number density.
However, this assumption is questionable wherever the magnetic field is strong enough to
affect the gas distribution. Assuming that the ISM is in pressure equilibrium, $n_e$ and $B$ are shown to be anti-correlated \citep{BSSW03}, which can significantly affect estimates of $B_\parallel$ from $\RM$. The relation between GMF strength and gas density in
the multi-phase ISM needs to be understood in detail in order to better
interpret the $\RM$ observations.

\begin{figure*}
$\begin{array}{cc}
\includegraphics[width=0.49\columnwidth,clip=true]{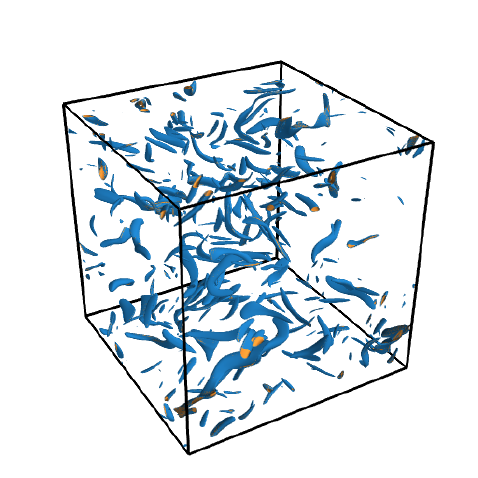} & 
\includegraphics[width=0.49\columnwidth,clip=true]{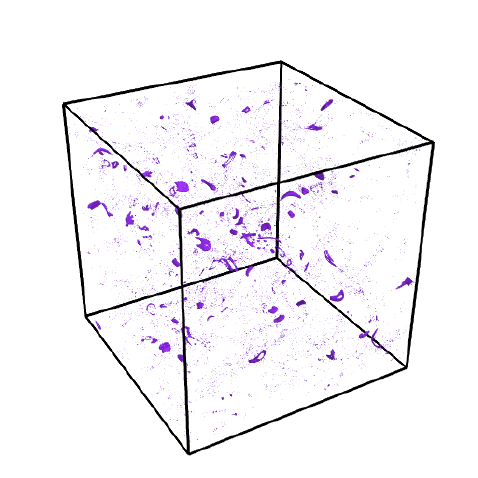} \\
\end{array}$
\caption{Test particle simulations of the propagation of cosmic rays in a random magnetic field
\jpr{(figures 2 and 8 in \citep{SSWBS18})}; the particles' Larmor radius is $6\%$ of the correlation length of the magnetic field. \textit{Left-hand panel:} isosurfaces of the strength of a random magnetic 
field $\vec{B}$ produced  by the fluctuation dynamo, $B^2/B_0^2=12$ (blue) and $15$ 
(yellow), with $B_0$ the root-mean square field strength. The magnetic field is intermittent, its correlation length is about $40$ times smaller than the correlation length of the chaotic flow that generates it.
\textit{Right-hand panel:} isosurfaces of the cosmic ray number density in the magnetic field of 
the left-hand panel at $n_\text{CR}/n_0=3.5$, where $n_0$ is the mean value of $n_\text{CR}$. The cosmic ray and magnetic field strength distributions are statistically 
independent. The number density of cosmic ray particles is larger in random magnetic traps between
magnetic mirrors.  Reproduced from \citep{SSWBS18}.  \textcopyright\
2017 The Authors Published by Oxford University Press on behalf of the Royal Astronomical Society.}
\label{CR-B2}
\end{figure*}

\paragraph{Synchrotron intensity}
\texttt{ }\\
Synchrotron radiation emitted by cosmic ray electrons interacting with
the GMF has an intensity $I(\nu)$ defined as
\begin{align*}
I(\nu) &\propto \int n_{\mathrm{CR}}(\nu, s)\, B_{\perp}^{1+\alpha}(s)\,\nu^{-\alpha}\, \dd s ,
\end{align*}
where $n_{\mathrm{CR}}(\nu,s)$ is the cosmic ray number density as a function of
frequency $\nu$ and path length $s$. Here, the cosmic ray number density per
energy interval $\dd E$ is assumed to follow a power law
\begin{align*}
n_{\mathrm{CR}}(E) &\propto E^{-\gamma}\dd E 
\end{align*}
with $\gamma = 2\alpha + 1$.
A typical value of the cosmic ray spectral index in the Galactic ISM
is $\gamma \approx 2.4$, which leads to the widely observed $\alpha =
0.7$. However, at low frequencies (typically below a few $100\,\MHz$), one
has to start taking into account effects like free-free absorption and
synchrotron self-absorption, which will alter the spectrum.
Since the synchrotron emissivity only
depends on the magnetic field component perpendicular to the line-of-sight, it is therefore complementary to the diagnostic of
Faraday rotation.

A significant complication in the interpretation of the observed synchrotron 
intensity is the uncertain knowledge of the number density of cosmic ray electrons. An assumption
almost universally adopted to estimate magnetic field strength from synchrotron intensity
is that of energy equipartition between cosmic rays and the GMF (or its variant), combined with a further 
assumption that cosmic ray electrons contain $1\%$ of the total cosmic ray energy or number density 
(\citealt{BK05}, and references therein). This assumption is applied to the synchrotron 
intensity observed locally, at the working resolution of the observations; in other words, 
$n_\mathrm{CR}$ and $|\vec{B}|^2$ are assumed to be perfectly correlated at all scales. 
Direct verifications of the CR--GMF equipartition based on gamma-ray observations are rare 
and inconclusive, and its physical justification is not entirely convincing. Analyses of the
synchrotron fluctuations in the Milky Way and M33 are not consistent with this assumption
at scales of order $100\,\pc$ \citep{Iacobelli:2013a,Stepanov:2014}. Using test particle 
simulations of cosmic ray propagation in random magnetic fields (either intermittent or Gaussian),
The distributions of $n_\mathrm{CR}$ and $|\vec{B}|^2$ 
are found to be not just uncorrelated but statistically independent \citep{SSWBS18}. Nevertheless, the two 
distributions are related, as shown in \ref{CR-B2}, but in a more subtle manner: cosmic rays 
are trapped between random magnetic mirrors whose occurrence is controlled not by magnetic 
field strength but rather by its geometry. These results strongly suggest that the CR--GMF
equipartition does not occur at the turbulent scales even though it may be 
relevant at $\kpc$ scales \citep{Stepanov:2014}. Understanding the relationship between the cosmic rays and the GMF requires
careful further analysis in order to convincingly interpret the synchrotron 
observations in terms of the magnetic field strength.

\paragraph{Synchrotron polarisation}
\texttt{ }\\
Synchrotron emission is intrinsically highly linearly polarised, with
a polarisation degree given by
\begin{align*}
\Pi &= \frac{3\gamma+3}{3\gamma+7},
\end{align*}
which means $\Pi = 72\%$ for $\gamma = 2.4$. However, this high \done{polarisation degree} is almost never observed, because the radiation is partially
depolarised when travelling from the source to the observer. The depolarisation
can be wavelength independent, due to small-scale tangling of the
magnetic field at the emission site, and/or wavelength dependent, due
to magnetic field tangling including Faraday rotation. For a
synchrotron emissivity $\epsilon(\vec{r}, \lambda)$, the observed
polarisation vector is
\begin{align*}
P\left(\lambda^2\right) &= \frac{\int\int \epsilon(\vec{r},\lambda) \Pi(\vec{r})\ee^{2\ii(\chi_0+\phi(\vec{r})\lambda^2)}\,\dd s\, \dd\Omega}{\int\int \epsilon(\vec{r},\lambda)
\,\dd s\, \dd\Omega } ,
\end{align*}
where integration over the solid angle $\dd\Omega$ defines the telescope beam.
The polarised synchrotron emission traced by the \planck\ mission at $30\,\GHz$ is shown in the top panel of \ref{fig:planck}.

\afterpage{
\addtocounter{footnote}{-1}
\begin{figure}[t]
\begin{centering}
\includegraphics[width=0.97\columnwidth]{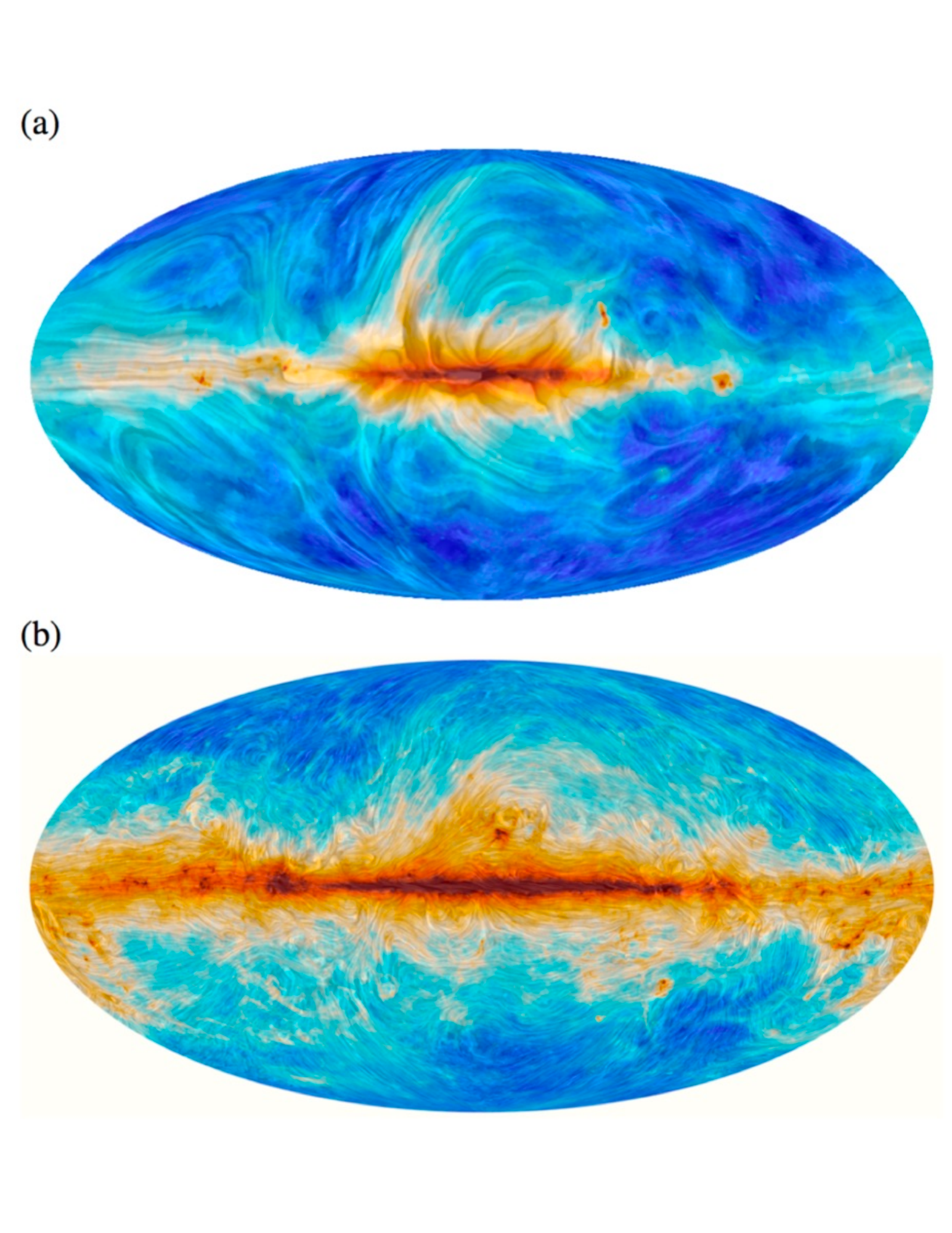}\end{centering}
 \caption[Synchrotron emission at $30\,\GHz$ and dust emission at $353\,\GHz$ ]{Synchrotron emission at $30\,\GHz$ ({\it top}) and dust emission at $353\,\GHz$ ({\it bottom}).
The colour indicates the total intensity, while the texture applied shows the inferred plane-of-sky magnetic field direction, \ie\ the polarisation direction rotated by $90\degree$.
See \citet{planck15_I} \jpr{for details.\footnotemark\ \sout{Image credit: ESA and the \planckC.}}}
\label{fig:planck}
\end{figure}
\footnotetext{\jpr{From \url{https://www.cosmos.esa.int/web/planck/picture-gallery}, reproduced with  permission from Astronomy \& Astrophysics, \textcopyright\  ESO;
original source ESA and the \planckC.
}}
}

\paragraph{Polarised dust emission}
\texttt{ }\\
The spin axis of a non-spherical dust grains is both perpendicular to its long axis and aligned, statistically, with the orientation of the local GMF \citep{andersson2015,hoang2016}.  Microwave, sub-millimetre, and far-infrared emission from these dust grains will therefore be polarised along the long axis of the grain, \ie\ perpendicular to the local magnetic field component projected onto the plane of the sky. The amount of alignment depends on local physical conditions and on 
the properties of dust grains (mainly their size and composition), but the analysis of \planck\ dust polarisation data indicates that the degree of grain alignment is high and homogeneous in the diffuse ISM including molecular cloud envelopes \citep{PIRXIX2015,PIRXX2015}. 
The polarised dust emission traced by the \planck\ mission at $353\,\GHz$ is shown in the bottom panel of \ref{fig:planck}.
The combination of \planck\ with higher frequency data from the BLASTPol ballon-borne experiment \citep{ashton2017} 
shows no dependence of the dust polarisation fraction $p$ on frequency.
This constraint, along with the ratio of polarised dust emission to the polarisation fraction of optical interstellar polarisation \citep{PIRXXI2015}, has been used to update dust models \citep{guillet2018}.
It suggests that the emission from a single grain type, which is efficiently aligned \citet{hoang2016}, dominates the long-wavelength emission of dust in both polarisation and total intensity.

\paragraph{Polarised optical/IR absorption by dust}
\texttt{ }\\
The same dust grains that emit linearly polarised emission absorb the optical and near infrared light from stars behind, causing a linear polarisation of the starlight parallel to the local magnetic field component projected onto the plane of the sky. This linear polarisation, like the dust emission polarisation, depends on the physical properties of the dust grains. 
Stellar polarisation data have been compared with sub-millimetre dust polarisation measured by \planck\ on the same line-of-sight. The two polarisation measurements have comparable sensitivity and are closely correlated \citet{PIRXXI2015,soler2016}. They offer complementary means to map the GMF, and to study its correlation with the structure and dynamics of interstellar matter. 
Stellar data are limited to a discrete set of sight-lines but they offer unique 3D information on the GMF 
using stellar distances. Optical polarisation data (e.g., \citealt{heiles2000}) are best suited to map the GMF over the diffuse ISM in the Solar neighbourhood \citep{berdyugin2014}, while near-IR polarimetric observations (e.g., \citealt{clemens2012}) probe the GMF within molecular clouds \citep{chapman2011} and the Galactic plane \citep{paveletal2012}.

\afterpage{\FloatBarrier}

\subsubsection{Large-scale magnetic field components}

The observable tracers summarised above make it convenient to describe the magnetic field as consisting of three components: a large-scale coherent or mean-field component (sometimes ambiguously referred to as the regular component); a small-scale random or turbulent component whose statistics are isotropic; and a third component that changes direction stochastically on small scales but whose orientation remains aligned over large scales, variously referred to as the anisotropic random component, the ordered random component, or the striated component. A description of how various observables trace these three components can be found in \citet{jaffe10}. First, we discuss the coherent component.

As in external spiral galaxies, the  magnetic field in the Milky Way is predominantly concentrated in the Galactic plane and approximately follows the spiral arms, as confirmed directly by starlight polarisation measurements and indirectly by modelling synchrotron radiation and RMs. A magnetic field reversal on scales of $\kpc$ has been unambiguously detected from pulsar RMs. However, observations cannot determine yet whether this reversal is azimuthal or follows the spiral arms, or whether this is a local feature or global along the entire spiral arm.

In current GMF models, spiral arms are commonly modelled as logarithmic spirals. However, there are multiple, heterogeneous, indications that this picture is oversimplified. GMF modelling of log-spiral arms independently in separate regions of the Galactic disk show that spiral arm locations and pitch angles seem to vary \citep{VBS2011}. The pitch angle of the magnetic field towards the anti-centre seems to be around zero \citep{RB2010}. In addition, in nearby face-on spiral M51, spiral arm pitch angles are variable with radius and azimuth and depend on their tracer: gas, dust or magnetic field \citep{PFS2006}. Also in the Milky Way, there are indications that GMF models with off-set dust and magnetic field spiral arms give better fits to synchrotron and thermal dust emission data \citep{Jaffe:2013}.

The difficulties associated with the complexity of
the observed spiral patterns, in either gas or magnetic
field, are compounded by the lack of a comprehensive theory
for the interaction of galactic dynamos with the spiral
pattern and the incompleteness of the theory of the spiral
patterns themselves \citep{CSQ14,CSS15}.
In particular, the 3D tracer of starlight polarisation will be imperative to constrain meso-scale deviations from logarithmic spirals, arm off-sets and/or varying pitch angles.

\afterpage{
\addtocounter{footnote}{-1}
\begin{figure}[tb]
\begin{centering}
\includegraphics[width=\columnwidth]{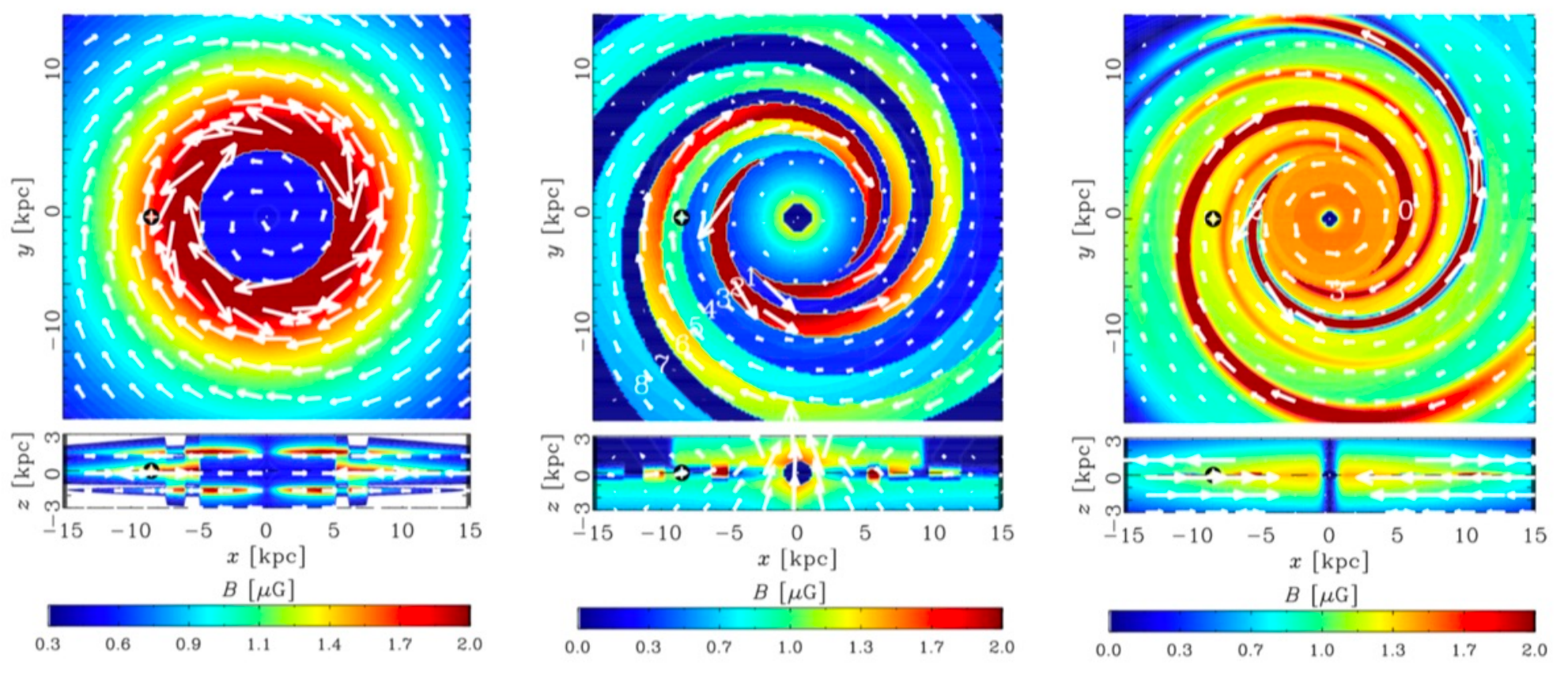}
\end{centering}
\caption[Three example models for the coherent magnetic field component in the Milky Way]{\jpr{Graphical representation\footnotemark\ of three} example models for the coherent magnetic field component in the Milky Way: on the {\it left} is from Sun et~ al. \citet{sun08}; in the {\it middle} is from Jansson \& Farrar
  \citet{jansson12b}; on the {\it right} is Jaffe et~al. \citet{Jaffe:2013} . The 
colour represents the strength of the coherent magnetic component, the white arrows show its direction. The top panel of each shows a cut through the Galactic plane at $z=0$ with the Sun position marked by the black plus, while the bottom panel of each shows a vertical cut intersecting the Sun and the Galactic centre. }
\label{fig:bcoh}
\end{figure}
\footnotetext{\jpr{Excerpt from figure~3 in Planck Collaboration, A\&A, Vol.~596, A103 (2016), reproduced with permission \textcopyright\ ESO.}}
}

The Milky Way has an out-of-plane magnetic field component, probably similar to X-shaped halo fields detected in nearby edge-on spirals. Its direction in the Solar neighbourhood is still debated \citep{hanqiao1994, maoetal2010} partly due  to insufficient data but primarily due to uncertainties in local structures in the Northern Galactic hemisphere.

The strength of the coherent magnetic field as derived from pulsar RMs is ${\sim} 2\,\muG$ (e.g., \citealt{hanetal2006}), generally consistent with modelling of synchrotron radiation. The total magnetic field strength is ${\sim} 6\,\muG$ at the Solar radius \citep{strongetal2000}, increasing to ${\sim} 10\,\muG$ at a Galacto-centric radius of $3\,\kpc$ \citep{beck2001}. Both the strength of the halo field and its scale height are very uncertain, with estimates in the literature of strengths between $2\,\muG$ and $12\,\muG$ and scale heights of ${\sim} 1.5\,\kpc$ from pulsar RMs, to $5\texttt{-}6\,\kpc$ from synchrotron emissivities, assuming equipartition with cosmic rays (for an overview, see \citealt{haverkorn2015}).

However, comparison with nearby spirals makes clear that many uncertainties remain in our knowledge of the large-scale magnetic field in the Milky Way: a variety of symmetries and dynamo modes are observed (which are as yet unobservable in the Milky Way); magnetic fields follow spiral arms in general but not in detail \citep{PFS2006}; the behaviour close to the Galactic centre (e.g., Galactic winds) is not understood; and large-scale reversals are not yet convincingly observed in nearby spirals (see \citealt{Beck16} for a review). Three example models from the literature are shown in \ref{fig:bcoh}; all were fit to synchrotron emission and Faraday RM data but clearly have very different morphologies, which illustrates the modelling challenge.

\subsubsection{Small-scale magnetic field components and interstellar turbulence}
\label{sss:turbulence}

One of the main challenges of ISM research is to allow for
the interconnectivity of all its components and their
interactions. Since the energy densities of the cosmic rays, turbulent
gas, radiation, and magnetic fields are comparable \citep{C2005, HH2012}, there are no passive tracers; instead, all ISM components deliver significant feedback to the system. 
Several studies show to what
degree these components are connected. The \planck\ dust polarisation maps 
showed that filaments in the diffuse cold neutral medium are
predominantly aligned with the local magnetic field \citet{PIRXXXII2016}, whereas filaments
in dense molecular clouds are mostly perpendicular to the
field \citep{PIRXXXV2016}. 
This correlation between the structure of matter and that of the GMF is also observed 
for filamentary structures identified in spectroscopic H\textsc{i} data cubes 
\citep{mccluregriffithsetal2006, clarketal2014}. More surprisingly, in some -- but
not all -- observations, Faraday depth filaments of polarised synchrotron emission seem to be aligned
with the local magnetic field in the dusty neutral medium
\citep{ZJD2015} or with H\textsc{i} filaments \citep{kalberla2016, kalberla2017}.

For decades, the power spectrum of the small-scale component of the magnetic field was approximated by a power law with a certain coherence length, assuming Gaussianity. Observational estimates for the coherence length vary: averages of extragalactic source RMs over large parts of the sky suggest coherence scales of about $100\,\pc$, while detailed studies in the Galactic plane suggest maximum scales of order $1\texttt{-}10\,\pc$ in spiral arms \citep{Haverkorn:2008} and in the highly-polarised Fan Region \citep{Iacobelli:2013a}.

However, the properties of magnetised interstellar turbulence \textit{cannot} be captured
using power spectra (and similar measures), because they are produced by random
variations in the components of the ISM that are not Gaussian. \done{A first phenomenological attempt to improve on a simple power law to denote the small-scale magnetic field in numerical simulations is to introduce two components: an isotropic random component generated by turbulence in the ionized gas, and an anisotropic random\footnote{also called `ordered' or `striated' in the literature} component, which is random in direction but uniform in orientation. This anisotropic random component can be created, e.g., by a shock wave compressing an isotropic random field, or by Galactic shear. This division is motivated by the observational tracers, as described in the previous section, and is a first approximation of the predicted non-Gaussianity of magnetised turbulence 
\done{(see also \ref{sss:MHDturb}).}
Global models show that the strengths of these random components are slightly higher than that of the coherent component but that the ratio of isotropic and anisotropic random fields varies across the Milky Way \citep{jaffe10,jansson12b,Jaffe:2013}.}

\begin{figure}[b]
\centering
\includegraphics[width=0.8\columnwidth,height=0.42\textheight]{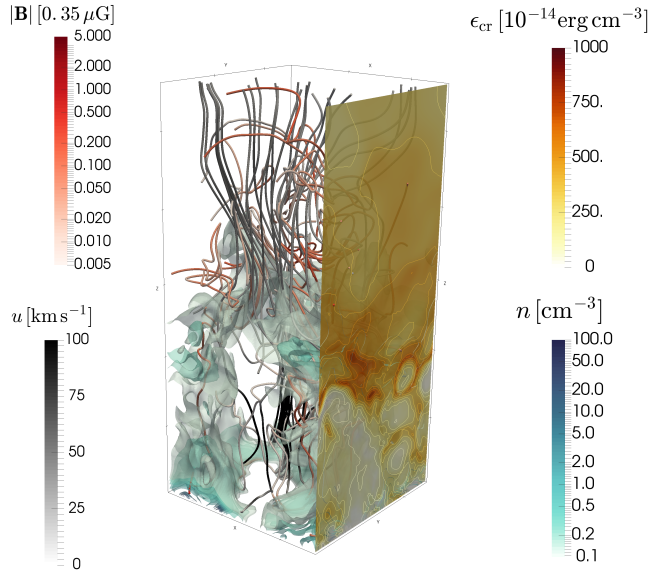}
\caption{\label{CR_box}Isosurfaces of gas density $n$ (blue), magnetic field $\vec{B}$ lines (red), gas 
velocity $\vec{u}$ streamlines (black) and cosmic ray energy density $\epsilon_{\mathrm{cr}}$ (contours 
on the right-hand face), in MHD simulations of the supernova-driven, multi-phase ISM 
that extend the simulations of 
Gent et\,al. \cite{Gent:2013b} 
by the inclusion 
of cosmic rays. Holes in the density distributions are supernova remnants; 
$z=0$ at the \textit{bottom}, $z\approx 2.2\,\kpc$ at the \textit{top}. 
(Simulation data: courtesy of G.~R.~Sarson, Newcastle University, UK.)}
\end{figure}

For this reason, synthetic observables that use Gaussian random fields to simulate fluctuations in the ISM do
not look the same as those generated from numerical simulations. Tools to
characterise shape and connectivity should be based on the mathematical theories of
morphology and topology. Measures using rigorously defined properties such as the
Minkowski functionals of morphology and the Betti numbers of topology, allow for
comparisons between different datasets (both observed and simulated) that better take
into account similarities and differences in their structure across different scales.
Applications of such measures to the ISM and MHD simulations have already been made
in morphology (e.g., \citealt{WBS07,MFS15}) and
topology (e.g., \citealt{Kowal:2007,Chepurnov:2008,Burkhart:2012,Makarenko:2017}).
A recent statistical study \citep{Henderson:2017} identified which combinations
of topological measures have the strongest discriminatory power, illustrating new,
effective ways in which these techniques can be used in practice. Methods from
mathematical morphology and topology can be incorporated into the \imagineSW\ framework to
allow for a more thorough and physically meaningful comparison between the different
real and simulated data related to the GMF and the ISM. The complexity of modern simulations of the supernova-driven ISM that include cosmic rays and allow for the mean-field dynamo action is illustrated in 
\ref{CR_box}.

\subsection{Dynamo theory as applied to the GMF}
\label{ss:DTGMF}

Turbulent dynamo
theory appears to be successful in explaining both the origin and
the basic observed structure of the GMF at both galactic and turbulent scales, \ie\ the mean ('coherent' or `large-scale') and random (fluctuating or `small-scale') magnetic fields.  The discussion in this section relies on a combination of insights from observations
and dynamo theory wherever they agree and, more importantly, where they do not.
A condensed but systematic
presentation of the current understanding of the galactic mean-field dynamo theory, with a compendium of useful qualitative results, can be found in \citet{CSSS14}. Here, we provide a brief summary.

The generation of the small-scale magnetic fields by the fluctuation
(or small-scale) dynamo relies on the random nature of
the interstellar gas flows at scales of the order of ${\lesssim} 100\,\pc$. If the electrical
conductivity of the plasma is sufficiently high,
so that the magnetic Reynolds number based on the flow correlation length exceeds a critical value of order $10^2$, a random plasma flow maintains a random magnetic field at scales smaller than the correlation scale of the flow
(\citealt{ZRS90,BS05}, and references therein). 
According to dynamo theory, large-scale magnetic fields arise spontaneously because of the overall galactic rotation and stratification of the turbulent ISM.
In the galactic context, they can be isolated in the total, partially ordered magnetic
field via spatial averaging, hence for this theoretical section, we will refer to them as the mean-field component.

\subsubsection{Mean-field dynamo}

The mean-field dynamo equations that govern the mean magnetic field
$\vec{B}$ can be given in the following simplest form:
\begin{align}
\label{MFDeq}
\deriv{\vec{B}}{t}=\nabla\times\left(\vec{V}\times\vec{B}+\alpha\vec{B}\right)
-\eta_\mathrm{t}\nabla^2\vec{B}\,,
\qquad \nabla\cdot\vec{B}=0\,,
\end{align}
where $\vec{V}=\left(V_r, r\Omega(r,z), V_z\right)$ is the large-scale (mean)
velocity field written in the cylindrical polar coordinates
$(r,\phi,z)$ with the origin at the galactic centre and the $z$-axis
aligned with the angular velocity $\vec{\Omega}$. The factor $\alpha$ is a
measure of the deviations of the interstellar turbulence from
mirror symmetry,
\[
\alpha\simeq\frac{l_0^2\Omega}{h}\,,
\]
with $l_0$ the turbulent correlation length and $h$ the scale height
of the layer that hosts the dynamo, and
$\eta_{\mathrm{t}} \simeq\frac{1}{3} l_0v_0\simeq 10^{26}\,\cm^2\s^{-1}$
is
the turbulent magnetic diffusivity with $v_0$ the root-mean-square (RMS) turbulent
velocity. This equation can be rewritten to allow for anisotropy
and inhomogeneity of the turbulence.

The single most important feature of a spiral galaxy that defines
the form of its large-scale magnetic field is the thinness of its gaseous disc, which constrains the mean
magnetic field to be oriented predominantly in the
plane of the disc.
Strong galactic differential rotation enhances
the azimuthal component $B_\phi$ at the expense of the radial one $B_r$.
As a result, the horizontal magnetic field assumes
the shape of a spiral with a relatively small pitch angle, typically
$p_B=\arctan\left(B_r/B_\phi\right)\simeq-(10\degree\texttt{-}20\degree)$.
The vertical magnetic field is weaker than the horizontal
components, but only \textit{on average}. Locally, and most importantly
near radial reversals of the magnetic field direction where
$B_\phi=B_r=0$, the mean magnetic field should be dominated by $B_z$.
Another location where $B_z$ is expected to dominate is a region within
about $1\,\kpc$ of the galactic centre.

Furthermore, because galactic discs are thin, the global structure of the mean
magnetic field within the disc is \textit{quadrupolar} in its spatial
symmetry: the horizontal components of the mean magnetic field have
the same sign above and below the mid-plane $z=0$, whereas the vertical
component is antisymmetric:
$B_{r,\phi}(-z)=B_{r,\phi}(z)$ and $B_z(-z)=-B_z(z)$.
Mean-field dynamo theory predicts that (quasi-)spherical bodies, unlike
thin discs, generate mean fields
of a \textit{dipolar} parity, where
$B_{r,\phi}(-z)=-B_{r,\phi}(z)$ and $B_z(-z)=B_z(z)$.
Such magnetic fields are observed in the Sun and the Earth. It is plausible that the mean-field dynamo operates in galactic haloes. If the two parts of the
halo are weakly magnetically connected, the mean magnetic field in the
halo can have dipolar symmetry, opposite to that in the
disc. Otherwise, the symmetry of the halo field may
be controlled by the disc. Unfortunately, detailed
and systematic observational information on the overall
symmetry of the large-scale magnetic fields in galactic
haloes is still lacking, while theoretical models remain
oversimplified and do not allow for definite predictions.

Interstellar magnetic fields are induced by
interstellar gas flows, and random (turbulent) flows are at
the heart of both small-scale and large-scale dynamos. It is
therefore not surprising that magnetic energy density scales
with the kinetic energy density of interstellar turbulence,
and the convenient measure of the magnetic field strength is that
corresponding to this equipartition,
$B_0=(4\pi\rho v_0^2)^{\nicefrac{1}{2}}\simeq 5\,\muG$,
where $\rho\simeq 1.7\cdot10^{-24}\,\g/\cm^3$ and $v_0\simeq 10\,\kms$
are the gas mass density and the RMS random velocity, respectively.

\subsubsection{Turbulent magnetic fields}
\label{sss:MHDturb}

Random magnetic fields in the ISM are produced and shaped by three
dominant processes: the tangling of the mean field by the random flows,
the fluctuation dynamo action and the compression by random shocks.
The latter two mechanisms produce spatially intermittent magnetic
fields where intense magnetic filaments, ribbons and sheets occur
in the background of weaker fluctuations (e.g., \citealt{WBS07,momferratos2014}).
Such magnetic fields have strongly non-Gaussian statistical properties
\citep{SSSBW17}. If the turbulent velocity has Gaussian statistics, so
does the magnetic field produced by the tangling of the large-scale field.
Estimates of the RMS magnetic field strength $b$ from these mechanisms
are rather vague, and $b\simeq B_0$ is the best available option.

\subsubsection{Beyond the basic theory}

The large-scale magnetic field produced during the kinematic (or exponentially growing) phase of a mean-field galactic dynamo is predicted to have rather specific symmetry properties. In particular, field patterns that are symmetric with respect to rotation about the axis of galactic rotation and that are also symmetric with respect to reflection in the mid-plane of the disc are predicted to grow fastest~\citep{Ruzmaikin:1988}. This preference for axial symmetry is supported by observations of a dozen nearby galaxies~\citep{Fletcher:2010}. However, the well established presence of a reversal in the direction of the Milky Way's mean magnetic field complicates the picture; the extent of the problem for the theory critically depends on whether the reversal is local (e.g., \citealt{Sh2005,VBS2011}) or
global (e.g., \citealt{hanetal2006}). As well as providing the best possible answer to this question given the available data, \imagine\ 
will also supply robust information about the strength and symmetry of the vertical component of the mean field. 

The symmetry properties discussed above are only strictly relevant during the kinematic phase of the dynamo. Once the field is amplified sufficiently it may become strong enough to modify the flow (at least its turbulent component) and the resulting non-linear feedback may affect the observed properties of the field. Alternatively, the saturation process may involve the transport of the magnetic field out of the Galactic disc by a wind or fountain flow \citep{Shukurov:2006}. Recent attempts to use observations of nearby galaxies to identify the saturation mechanism were inconclusive \citep{VanEck:2015}, possibly because the available observables, in particular the magnetic pitch angle, are not sufficiently sensitive. A more detailed determination of the properties of the Milky Way's magnetic field, which will not be possible for other galaxies in the medium term, could help to solve the problem. In particular, recent results from numerical simulations \citep{EGSFB16,EGSFB17} show that the saturated mean-field can have a maximum that is displaced from the mid-plane by a few hundred parsecs and that the ISM phase structure is modified as the field saturates. \imagine\
offers the prospect of determining the vertical profile of the magnetic field and thus makes a vital connection between observations and simulations of a mean-field dynamo in its non-linear state;  this will enable the theory to move to a more advanced level.

The dynamo theory in its present form largely neglects the multi-phase structure of the ISM. Numerical simulations suggest that the large-scale magnetic field is maintained in the warm gas whereas the turbulent field is less sensitive to the multi-phase structure \citep{EGSFB16}. It remains unclear whether the mean-field dynamo acts within the complex and erratically evolving volume occupied by the warm phase alone or whether it responds to an `average' ISM whose parameters can be obtained by spatial and temporal averaging. This problem can only be solved by the coordinated observational, theoretical and numerical efforts envisaged by \imagine. 

\subsubsection{Plasma processes} 
\label{PP}

The spectrum of interstellar magnetic fields is known to
extend from scales of order $100\,\pc$ down to Earth diameter scales (${\sim} 10^{9}\,\cm$) and less \citep{ARS95,ChLa10}. 
It is argued in \citet{SCDHHQT09} that
the fluctuations extend down to the ion and
electron Larmor radii where the turbulent energy is
converted into heat. These are of order $10^{8}\,\cm$
and $3\cdot10^{6}\,\cm$, respectively, in the warm ionized
interstellar gas. The dynamics of the turbulence in the inertial
ranges, formed by non-linear interactions that vary widely
in their physical nature across this broad range of scales,
is universal and independent of the properties of the magnetic
field at the larger scales of order $100\,\pc$ where the turbulence is driven.
However, the energy input into the turbulent cascade is
controlled by the properties of the GMF at scales larger
than $100\,\pc$. Moreover, intermittency can destroy the
scale invariance of the turbulent cascade \citep{CSM15}.

Essentially, interstellar turbulence is compressible,
intermittent, MHD
turbulence whose nature is complicated by the multi-phase
ISM structure that leads to vastly different physical
conditions at different positions and, at a fixed position,
at different times. MHD turbulence theory has lately
been advanced significantly
(\citealt{SG94,GS95,BL05,BL06,Bol06,MCB06,LGS07}, and references therein), albeit not
without controversy. Further insights into the nature and
significance of the ISM turbulence are likely to follow
from a recent extension of the theory of magnetised
turbulence from an MHD approximation to kinetic plasma
theory approaches \citep{SCDHHQT09,BHXP13,BCXZ15}.
The astrophysical significance of the theoretical
progress
remains to be understood, and
appropriate observational techniques are waiting to be
fully developed and implemented. The forthcoming
observations with next-generation radio telescopes should
permit a critical assessment of the theory.
Apart from insights into the large-scale magnetic fields, \imagine\ 
would advance our understanding of magnetic fields at the turbulent scales and reveal the magnetohydrodynamic and plasma processes that control them.

Neither the MHD theory nor the observational modelling
alone will make much progress on Galactic magnetic fields. \imagine\ 
will join the two in order to leverage their combined power and 
meet this important astrophysical challenge.

\subsection{Indirect dark matter detection} 
\label{ss:DM}

We know that dark matter is the main driving force in structure formation, and thus crucial
for the creation and evolution of galaxies such as our own, but we have not pinpointed
its identity yet. There are three complementary ways of elucidating the nature of
the DM particle and its properties \citep{DMind}: (1) the production of DM in accelerator
experiments; (2) direct detection by observing nuclear recoils from interactions of DM in a detector; 
and (3) indirect detection, \ie\ 
experiments that search for stable
secondaries, 
produced by annihilation or decay, that accumulate, e.g., in
the Milky Way or its surrounding dwarf galaxies. 
In this last approach, experimental efforts focus mostly on neutral messengers such as gamma-rays or neutrinos, which are particularly relevant to exploring the parameter space of heavy (${\sim} \PeV$) dark matter, and which are investigated with 
state-of-the-art or upcoming instruments like H.E.S.S., VERITAS, MAGIC, and CTA for 
gamma-rays as well as IceCube and KM3NeT for neutrinos. 

\done{A relevant charged particle channel for indirect detection of DM would be an excess of antimatter} relative to the predictions for astrophysical backgrounds\done{, as matter and antimatter are produced in equal amounts in DM annihilations or decays.} Clearly, the identification of any \done{DM signal in, e.g., the positron fraction in GeV cosmic rays seen by PAMELA 
and AMS \cite{pam1, 2013PhRvL.110n1102A}}, requires
an adequate understanding of the corresponding backgrounds. This requires improved
modelling of the propagation of astrophysical cosmic rays, which in turn relies on a better
understanding of the GMF. Moreover, predicting the secondary fluxes from
DM annihilations or decays depends strongly on the properties of the GMF inside the
Galactic halo, because DM resides in an extended, approximately spherical halo.
Even in the case of a non-detection of DM, a more precise knowledge of the GMF would
allow one to improve limits on DM properties.

\section{Extragalactic science}
\label{s:BGEG}

The impact of the \imagine\ project reaches far beyond our own galaxy. Here we summarise the intimate connections between the goals of \imagine\ and three of the most active areas of extragalactic astrophysics and cosmology.

\subsection{Structure formation}
\label{ss:SF}

According to the current paradigm of cosmological structure formation all observable structures in the Universe originated from weak primordial quantum fluctuations generated during the early epoch of inflation. This model connects the \done{large-scale} dynamics of the Universe and the growth of inhomogeneous structures with an underlying gravitational world model. The currently favoured $\Lambda$CDM model assumes that the Universe is governed by general relativity and that its homogeneous, large scale dynamics can be described by a special case of a Friedmann-Lema\^{i}tre-Robertson-Walker metric. 

\done{Most of our current cosmological knowledge originates from observations of the homogeneous expansion of the Universe via observations of type Ia supernovae, primordial temperature fluctuations in the CMB, or the linear fluctuations of matter at the largest scales in galaxy observations (see, e.g., \citealt{2004mmu..symp..270F,2006astro.ph..1168L} and references therein).}
The $\Lambda$CDM model has been demonstrated to fit all these data, in particular the high precision CMB observations of the \planck\ mission to astonishing accuracy (see e.g., \citealt{2016A&A...594A..13P}). These results suggest 
that the gravitational evolution of the present Universe is governed by enigmatic dark matter and dark energy, constituting up to about $95\%$ of the total cosmic energy budget.
Although required to explain the formation of all observable structures within the standard picture of Einstein's gravity, so far dark matter and dark energy elude direct observations and have not yet been identified as particles within more fundamental theories \citep{2017IJMPD..2630012F}.

New challenges arise from studying the non-linear and inhomogeneous distribution of matter in our Universe in greater detail. Due to non-linear gravitational interactions dark matter aggregates into massive clusters and cosmic filaments forming the so-called cosmic web. This three-dimensional configuration of matter is believed to be a unique result of the primordial initial conditions, set by inflation, and the subsequent hierarchical structure formation history in a cold dark matter scenario. These filaments form the pathways along which matter accretes onto the massive galaxy clusters, as observed in cosmological surveys.
To study this filamentary distribution of matter in the Universe, novel Bayesian methods infer accurate and detailed maps of the 3D DM distribution from galaxy observations in the nearby Universe (see e.g., \citealt{2013MNRAS.432..894J,2015JCAP...01..036J,2016MNRAS.455.3169L}).
These methods fit 3D numerical models of non-linear gravitational structure formation to the data. In doing so they simultaneously reconstruct the initial conditions from which present cosmic structures originate as well as non-linear density and velocity fields in the present Universe including their respective dynamic formation histories (see e.g., \citealt{2016MNRAS.455.3169L}). This provides a completely new view of the dynamical large-scale structure formation in the nearby Universe and enables detailed studies of the non-linear DM distribution in nearby structures, such as the Coma or Shapley cluster. For the first time, these methods provide us with a conclusive statistical model of the large-scale 3D structure surrounding us and a detailed characterisation of its origins and non-linear formation histories. 

The understanding of the large-scale structure of our local Universe and its formation history affects the extragalactic science done within \imagine\ in multiple ways.  In particular, it serves as a fundamental prior for the structure of extragalactic magnetic fields and the distribution of UHECR sources and thus constrains the expected distribution of UHECRs outside the Galaxy. This information is required to use UHECRs as tracers for the large-scale GMF structure, as explained in \ref{sec:UHECR_tracer}.

\subsection{Galaxy formation and evolution}
\label{sec:gal_form}

One of the most important aspects of cosmological structure formation in the context of \imagine\ is that of the formation and evolution of galaxies, with an immediate focus on the growth and evolution of their magnetic fields. A similarly interesting aspect is the potential of \imagineSW\ to aid in understanding some of the physical processes underlying galaxy formation. There are two possible routes: (a) \imagineSW\ analyses of the GMF in the Milky Way that can probe (non-thermal) feedback processes at exquisite angular resolution, but  limited to low-to-moderate star formation rates in our galaxy; and (b) adapting \imagineSW-inspired tools to observations of nearby starburst galaxies, which in some respects resemble high-redshift galaxies that form stars at high rates at the peak of the galaxy formation epoch. \done{Conversely, data obtained from observations of other galaxies can be used in \imagineSW\ as prior information for modelling the GMF, as discussed in \ref{sss:Galaxies}.} 

Within the $\Lambda$CDM paradigm, many aspects of galaxy formation and evolution are not well understood and appear to be in conflict with the data. Most prominently, the observed galaxy luminosity and H{\sc i}-mass functions show much shallower faint-end slopes than predicted; this is locally known as the ``missing satellites problem'' of the Milky Way \citep{Klypin1999,Moore1999}. At the same time, simulations predict an inner DM density cusp in galaxies seemingly at odds with the cored profiles observed in low surface brightness galaxies and dwarf satellites. While these problems may point to an incomplete understanding of the underlying theory of DM (e.g., \citealt{vandenAarssen2012}), they also highlight our inadequate understanding of galaxy formation, particularly the effects of cosmic rays and magnetic fields.

It is believed that the problem can be resolved by inclusion of \textit{feedback} processes by stellar winds, supernovae, and active galactic nuclei, which either drive outflows of gas from the galaxy and/or interrupt the accretion of new gas via a self-regulated heating mechanism.  These processes lead to the suppression of star formation activity in both small mass halos ($M\lesssim 10^{11}\,\Msun$) and large mass halos ($M\gtrsim 10^{13}\,\Msun$). The physical processes underlying these ideas include:  energy and/or momentum input by cosmic rays; magnetic fields; radiation fields; and mechanical energy in the form of shock waves and turbulence. But details remain largely unclear. Numerical simulations of the present-day ISM (e.g., \citealt{WalchEtAl2015, GirichidisEtAl2016, Simpson2016, GattoEtAl2017}) indicate that the details of the density structures as well as the positioning of the supernovae are crucial to understand the dynamical cycle of gas in the ISM and the launching of outflows.  In recent years, cosmic ray feedback has been rediscovered and has attracted a growing amount of work \citep{UPSNES12,BAKG13,Hanasz2013,Salem2014,Girichidis2016,pakmoretal2016,Simpson2016}. 

The formation of galaxies follows two main phases: an early violent phase at $z\gtrsim1$ with large mass accretion rates and high gas-mass fractions, and a late epoch ($z\lesssim1$), which is characterised by lower gas accretion rates and some (rarer) galaxy merger events. Understanding the non-thermal components of the present-day Milky Way will not directly teach us about feedback processes that have been active at the high-noon of galaxy formation at $z\sim2$. However, the combination of non-thermal tracers from the most star-forming regions in the Milky Way with resolved observations of local starburst galaxies (such as Arp 220 and NGC 253) has the power to improve our understanding of the physical feedback processes at work during the galaxy formation epoch. Thus, understanding non-thermal components of the ISM via Bayesian inference methods such as \imagineSW\ promises to be an essential step towards better motivated feedback processes.

Modern models of galaxy formation are now starting to include MHD. The turbulent fluctuation dynamo can quickly (on a time-scale of order $10\,\Myr$) amplify a small-scale magnetic field, even in a very young galaxy. In later stages when galaxies grow large disks, the field is further amplified by a large-scale dynamo driven by differential rotation and turbulence, and this process preferentially grows a toroidal disk field \citep{pakmoretal2016, Pakmor2017, Pfrommer2017}. Numerical simulations show that magnetic fields can grow in about a $\Gyr$ to a saturation value where magnetic pressure is comparable to thermal pressure if turbulence is sufficiently strong \citep{wangabel2009, riederteyssier2016}. While this magnetic field can partially suppress and delay the onset of star formation \citep{PS2013}, overall, magnetic fields have only a weak effect on the global evolution of the galaxies \citep{Pakmor2017}. However, their presence is critical to correctly model cosmic ray feedback. This is because a strong starburst injects cosmic rays that accumulate to the point where their buoyancy force overcomes the magnetic tension of the dominant toroidal magnetic field, causing it to bend and open up \citep{Rodrigues2016}. Cosmic rays stream and diffuse ahead of the gas into the halo and accelerate the gas, thereby driving a strong galactic outflow \citep{pakmoretal2016,Pfrommer2017b}. 

These simulations predict a local magnetic field strength of a few $\muG$ in the galactic disc, increasing to $10\texttt{-}100\,\muG$ close to the galactic centre, in magnetically dominated galactic winds \citep{pakmoretal2016,Pfrommer2017b}, or in the X-shaped magnetic fields out of the disc \citep{hanaszetal2009}. To test those simulations against observables, we need to capture the characteristics of these meso-scale outflows in a GMF model, or even to develop a non-parametric reconstruction of the GMF that can avoid our morphological preconceptions for what they should look like. Here is where the advanced methodology of \imagineSW\ can have an impact (see \ref{ss:non-para-models}) and conversely, where \imagineSW\ can use this information about galaxy evolution for improvements in the modelling of the GMF (see also \ref{sss:Galaxies}).


\subsection{Ultra-high energy cosmic rays}
\label{ss:UHECR}

Cosmic rays with energies exceeding $E \sim 10^{18}\,\eV = 1\,\EeV$ are called ultra-high energy cosmic rays and are assumed to be mostly of extragalactic origin \citep{Kotera:2011cp,2018PrPNP..98...85M}. Though more than $20\,000$ events have been observed in the highest energy decade \citep{ThePierreAuger:2015rha,Jui:2016amg,Fenu:2017sdj},
the identification of their sources still requires the solution of three problems: (a) understanding the structure and strength of cosmic magnetic fields (both Galactic and extragalactic), which determine the path of the particles on their way from their sources to us; (b) clearly identifying the mass and charge of \textit{individual} UHECR particles, and thus the magnitude of their magnetic deflections; and (c) identifying statistically significant anisotropy in UHECR arrival directions, which is required to compare potential source distributions in the sky with measurements. In all three areas, significant experimental progress has been reported recently, which is laid out as follows.
\afterpage{
\addtocounter{footnote}{-1}
\begin{figure}[t]
\centering
\includegraphics[width=0.95\columnwidth]{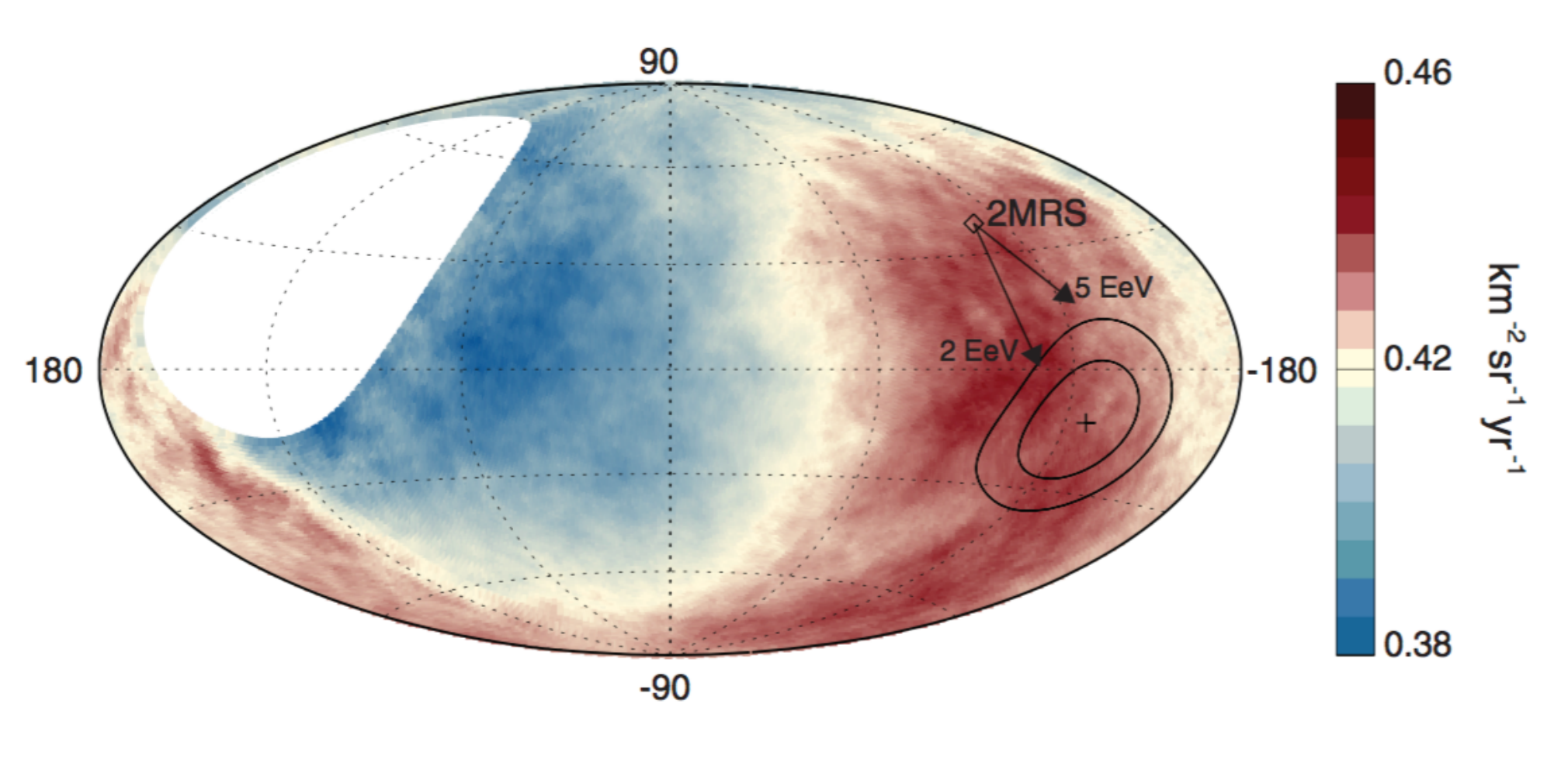}
 \caption[Sky map
in galactic coordinates showing the cosmic ray flux as measured by the Pierre Auger Observatory for $E>8\,\EeV$]{Sky map
in galactic coordinates showing the cosmic ray flux as measured by the Pierre Auger Observatory for $E>8\,\EeV$ smoothed with
a $45\degree$ top-hat function \jpr{\citep{2017Sci...357.1266P}}. The Galactic centre is at the origin. The cross
indicates the measured dipole direction; the contours denote the $68\%$ and $95\%$
confidence level regions. The dipole in the 2MRS galaxy distribution is
indicated. Arrows show the deflections expected for the JF12 GMF model on particles with $E/Z = 2\text{ or }5\,\EeV$.\footnotemark
\jpr{\sout{Image credit: Pierre Auger Collaboration \citep{2017Sci...357.1266P}}}. 
\label{auger-skymap}}
\end{figure}
\footnotetext{\jpr{Figure~3 in \textit{Observation of a large-scale anisotropy in the arrival directions of cosmic rays above $8\,{\times}\,10^{18}\,$eV}, Pierre Auger Collaboration, Science, Vol. 357, Issue 6357, pp. 1266-1270, \textcopyright\ (2017), reprinted with permission from AAAS.}}
}

While past attempts to associate UHECRs with known extragalactic structures \citep{Stanev:1995my, Abraham:2007si} could not be confirmed with sufficient significance, the analysis of a much larger dataset obtained by the southern Pierre Auger Observatory \jpr{\sout{(PAO)}} found a large-scale dipole anisotropy with an amplitude of $6.5\%$ for energies above $E=8\,\EeV$ and a significance of $5.2\sigma$ (\citealt{Aab:2016ban} and \ref{auger-skymap}), similar to what was found in the combined analysis with the northern Telescope Array (TA) experiment \citep{Aab:2014ila}. Moreover, a recent analysis of the cumulative dataset from the \jpr{\sout{PAO} Pierre Auger Observatory} provides evidence for anisotropy in the arrival directions of UHECRs on an intermediate angular scale, which is indicative of excess arrivals from strong, nearby sources \citep{AugerStarburst}, in line with previous results of the \jpr{\sout{PAO} Pierre Auger Observatory} \citep{PierreAuger:2014yba} and TA \citep{Abbasi:2014lda}. 

\afterpage{
\addtocounter{footnote}{-1}
\begin{figure}
\centering
 \includegraphics[width=0.93\columnwidth]{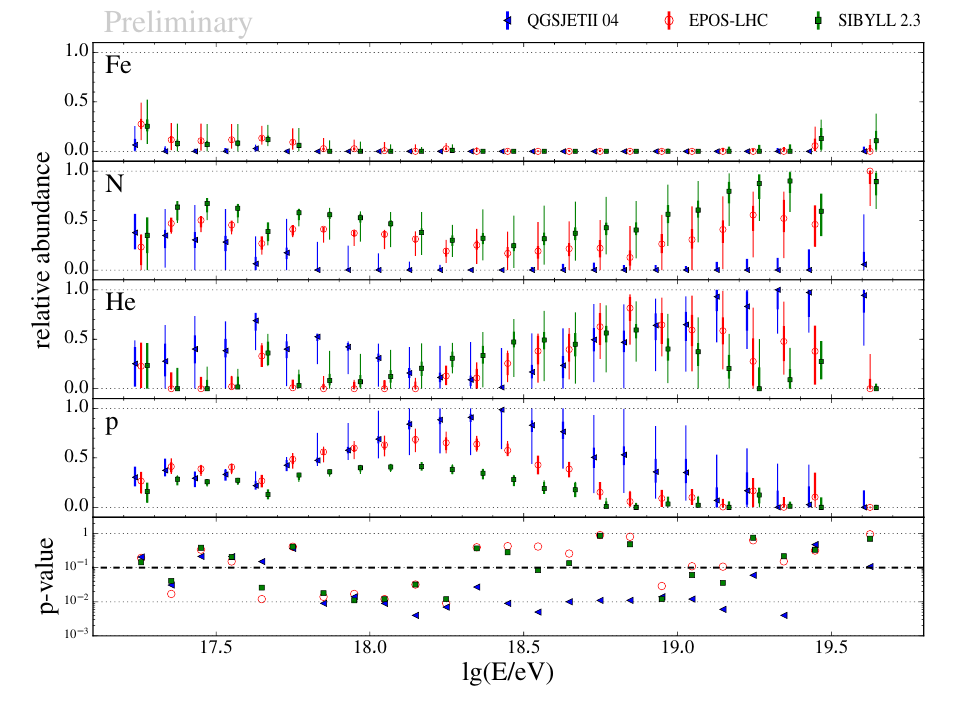}
 \caption[Relative abundance of four mass groups as function of energy in cosmic rays as measured by the Pierre Auger Observatory.]{Relative abundance of four mass groups as function of energy in cosmic rays as measured by the Pierre Auger Observatory \jpr{\citep{Bellido2017}}. The upper four panels show the best-fit mass fractions and the goodness of fit is displayed in the lowest panel. Thick error bars denote the statistical uncertainties, thin error bars the systematic ones.\footnotemark \jpr{\sout{Image credit: Pierre Auger Collaboration \citep{Bellido2017}.}} 
\label{auger-massspec}}
\end{figure}
\footnotetext{\jpr{Figure~6 in J.~Bellido et al., PoS (ICRC2017) 506, reproduced with permission \textcopyright\ by the Pierre Auger Collaboration.}}
}

The problem of identifying the charge of UHECR nuclei arises from the fact that it has to be inferred from the observed properties of air showers, cascades of secondary particles developing in the Earth's atmosphere initiated by the UHECR. While the energy is quite well constrained by shower calorimetry, the mass and charge of the primary particle reveals itself mostly by the atmospheric column depth of the maximum energy emission of the shower, $X_{\rm max}$, which is subject to large fluctuations between individual showers. This will be brought under control by collecting additional information on individual showers, and efforts are under way to achieve this,
such as the upgrade of the \jpr{\sout{PAO} Pierre Auger Observatory} \citep{Aab:2016vlz}, including a new method to measure air showers by radio detection \citep{Buitink:2016nkf, Schulz:2015mah}. Currently, the primary mass is only measured for a fraction of UHECRs. 
There is evidence that the \textit{average} composition of UHECR nuclei is rather light at ${\sim}\,1\,\EeV$ (\citealt{kampert2012,Thoudam:2016}, and references therein), becoming heavier towards the highest energies. Recent results from the \jpr{\sout{PAO} Pierre Auger Observatory} are illustrated in \ref{auger-massspec}, depicting the relative abundance of four elemental groups in cosmic rays. 

These results indicate that \textit{most} UHECRs have rigidities in the range $E/eZ \sim 3\texttt{-}10\,\EV$, which means that they would be deflected in the Galactic magnetic field by ${\gtrsim} 10\degree$, considering both the large-scale and the random component \citep{Giacinti:2011uj,Keivani:2014zja,Farrar:2017lhm}. This is in addition to the deflection in extragalactic magnetic fields, which is considerably more difficult to quantify but unlike the GMF, only \textit{diminishes} the directional information by blurring the UHECR sources over more or less large cones, an effect which can be controlled by statistics. The systematic shift of the centre of the cone away from the source can then be used to constrain \textit{both} GMF properties and UHECR source scenarios in the Bayesian analysis of \imagineSW. 

An improved understanding of the GMF structure would be the first step toward a more rigorous UHECR astronomy. Reducing the uncertainties between measured and extragalactic arrival directions of UHECR events  would allow for the comparison of different scenarios for the large-scale extragalactic structure and
distribution of UHECR sources and magnetic fields. This would not only make crucial progress
in identifying UHECR sources possible, 
but also allow us to constrain extragalactic magnetic fields (EGMFs) by deducing the residual extragalactic contribution to magnetic deflection once the Galactic contribution has been subtracted. Comparing this with constrained EGMF simulations (e.g., \citealt{2018MNRAS.475.2519H} and \ref{fig:egmf_Hackstein}) will contribute to the understanding of
the astrophysical processes relevant for MHD at large cosmological scales, such as large-scale dynamo processes
and the development of MHD turbulence at galaxy cluster scales and beyond. At present, many of such EGMF simulations exist (e.g., \citealt{PhysRevD.70.043007, 1475-7516-2005-01-009, 0004-637X-682-1-29, PhysRevD.77.023005, 2016MNRAS.462.3660H, 2018MNRAS.475.2519H}), varying largely in their predictions for magnetic field strengths on various scales and effects on UHECR propagation \citep{PhysRevD.96.023010}.

\subsection{Extragalactic backgrounds}
\label{ss:exbkg}

\subsubsection{Cosmic microwave background}
\label{ss:cmb}

Studies of the cosmic microwave background (CMB) and its anisotropies have ushered in a new, high-precision era for modern cosmology.
Whether alone or combined with other cosmological probes, measurements of the CMB total intensity anisotropies have established the current cosmological model, setting the stage for further, more profound investigations with direct implications not only for cosmology but also for fundamental physics.
Today, the search for primordial gravitational waves from the inflationary phase of the expanding Universe is the paramount goal of CMB experiments. The signal imprinted on the polarised  CMB, the so-called primordial $B$-modes\footnote{Cosmologists decompose the polarised emission into $E$ (gradient-like) and $B$ (curl-like) modes (e.g., \citealt{Caldwell16}). These correspond to signals of distinct physical origin within the polarisation of the CMB.}, are directly related to physics beyond the Standard Model of particle physics and on energy scales twelve orders of magnitude higher than those accessible to the Large Hadron Collider.

\done{The power spectra analysis of the \planck\ $353\,\GHz$ polarisation maps decomposes the dust polarisation into $E$ and $B$ modes \citep{PIRXXX2016}. It led to two unexpected results: a positive $TE$ correlation and a ratio of about $2$ between the $E$ and $B$ dust powers. More recently, the $TE$ correlation and $E/B$ power asymmetry were shown to extend to low multi-poles that were not analysed in the first \planck\ polarisation papers \citep{PIRLIV2018}. This latter study 
also reports evidence for a positive $TB$ dust signal. The 
$E/B$ asymmetry and the $TE$ correlation have been related empirically to the alignment of the magnetic field with the filamentary structure of the diffuse ISM \citep{planck2015-XXXVIII}. Theoretically, these results have been interpreted as signatures 
of turbulence in the magnetised ISM \citep{Caldwell16,kandel18,Kritsuk17}.
The $TB$ signal indicates that dust polarisation maps do not satisfy parity invariance, a surprising 
result that has not yet been explained. }

After the \planck\ mission, a new generation of experiments, on the ground and balloon-borne, is measuring the polarisation of the CMB with an increased precision. The primordial $B$-mode signal may well be within reach of these experiments' sensitivities, but it is much weaker than the polarised foreground emission from the magnetised ISM of our galaxy.
What we learn in modelling the GMF and its interactions with the various components of the ISM will be a useful input to the CMB component separation challenge. A better understanding of the polarised emission from dust and the relationship between the field and the dust emission structures will lead to more realistic foreground simulations for testing component separation algorithms. It is also vital in order to compute realistic errors and eventually to claim a detection of the cosmological signal with confidence \citep{vansyngel2017}. At low frequencies, a better characterisation of the synchrotron emission would be crucial to model properly the anomalous microwave emission for component separation purposes.
Furthermore, detailed knowledge of these Galactic emission mechanisms at large scales are also a key for detecting other sources of $B$-modes, such as those coming from primordial magnetic fields \citep{planck19} (see also \ref{sss:primordialB}).

The CMB measurements and the detection of primordial gravitational waves are therefore intimately linked to the inference of the GMF: the measurements in the microwave bands are crucial inputs to the \imagineSW\ framework, and the resulting knowledge of the thermal and non-thermal components of the magnetised ISM is a crucial input to the challenge of cosmological component separation.

\subsubsection{Epoch of reionisation}

One of the greatest outstanding challenges of observational astronomy at the moment is the detection of neutral hydrogen from the era when the first galaxies started to form and reionise the Universe, the so-called epoch of reionisation (EoR). The $21\texttt{-}\cm$ neutral hydrogen line will be redshifted to low radio frequencies, where it is expected to be detectable statistically with the current generation of low-frequency radio telescopes \citep{asadetal2017}. However, as in the case of the CMB, the signal will be much weaker than several foreground components, so extreme care needs to be taken to remove these.

One of the most difficult foregrounds to remove may be spurious small-scale spectral structure caused by polarisation leakage (e.g., \citealt{jelicetal2008}). If ideal radio telescopes existed, EoR measurements would not be connected in any way with the Galactic magnetic field. However, in reality, leakage of diffuse polarised synchrotron emission from the Milky Way into total intensity will result in a frequency-dependent signal in total intensity that mimics the EoR signal itself. If the polarised foreground is known, this leakage in principle can be calculated as a power spectrum and its contribution subtracted from the observed signal \citep{asadetal2017}.

\begin{figure}[tb]
\centering
\includegraphics[width=0.98\columnwidth]{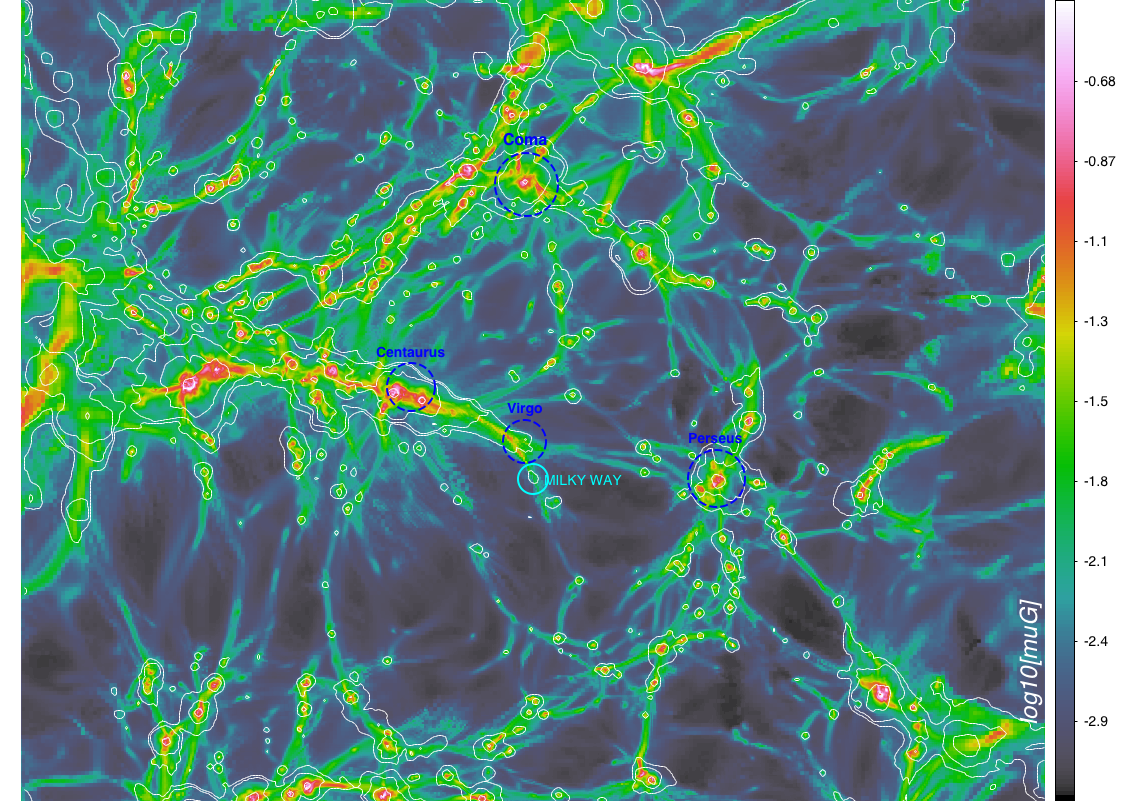}
\caption{
Map of projected mean magnetic field along the line-of-sight obtained from an MHD simulation of the local Universe as used by Hackstein et al. \citet{2018MNRAS.475.2519H}. The simulation started from constrained initial conditions provided by a method summarised in \citet{2016MNRAS.455.2078S}. The catalogue of constraints is fully described in \citet{2013AJ....146...86T}. The magnetic field is shown in $\muG$. Colours are in logarithmic scale. The additional contours show the temperature in logarithmic scale. The panel has a side-length of $200\,\Mpc/\mathrm{h}$, the projection axes are the $X$ and $Y$ in the super-galactic coordinates. The position of the Milky Way is at the centre of the box, indicated by a white circle. The additional circles show the location of the simulated counterparts of real objects in the local Universe. \label{fig:egmf_Hackstein}}
\end{figure}

\subsubsection{Large-scale extragalactic and primordial magnetic fields}
\label{sss:primordialB}

\begin{figure}[t]
\centering
\includegraphics[width=\columnwidth]{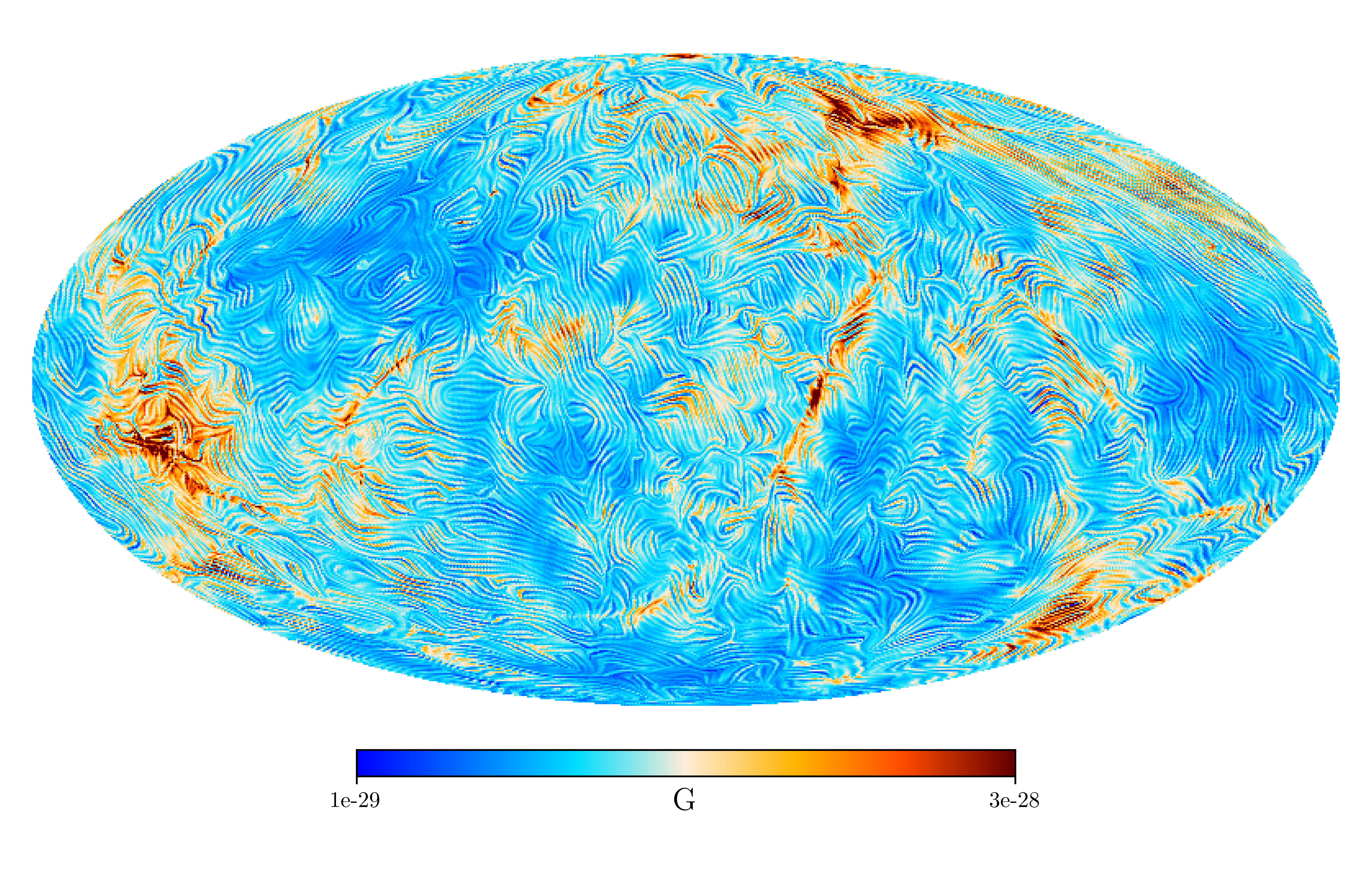}
 \caption{All-sky map of the extragalactic primordial magnetic field as it should have been generated by the Harrison mechanism in the early Universe and be still present today. Shown is the field strength in colour and the field direction as a texture, where each pixel represents the average over the line of sight to a distance of $60\,\Mpc/\mathrm{h}$ from the Earth. 
The 3D field was  reconstructed from the observed galaxy distribution in  the local Universe \citep{2018arXiv180302629H}. Galaxy clusters (like Virgo on the top right) with compressed primordial fields appear in red, galaxy voids with mostly pristine fields in blue.}
\label{fig:PMF}
\end{figure}

In recent years, a large effort has been devoted to study cosmic
magnetism (e.g., \citealt{Johnston-Hollitt2015}), but numerous
questions about the origin and evolution of magnetic fields in the
Universe are still unanswered. Important information could be derived from
the observation of magnetic fields in the large-scale structure of the
Universe, beyond galaxy clusters. In these environments, the properties of
the magnetic fields reflect those of the seed field, since they have been less affected by processes of structure formation (see \ref{fig:PMF}). But also magnetic fields in filamentary structures of the cosmic web are still expected to be very weak ($B\sim 1\,\nG$, see e.g., \citealt{Vazza2014}). The detection of magnetic fields outside galaxy clusters is therefore challenging, but would prove invaluable in understanding the origin of primordial fields.

As pointed out in \ref{ss:cmb}, an imprint of these primordial magnetic fields must be present in the CMB. The current experimental upper limit on primordial magnetic fields from CMB observations is $B < 1\,\nG$ (for a nearly scale-invariant stochastic distribution), but most theoretical scenarios to produce primordial fields expect much smaller field strengths (\citealt{planck19}, and references therein). 
A mechanism to measure such weak fields has been suggested on the basis of pair cascades that are expected to develop in the traversal of $\TeV$ photons through cosmic radiation backgrounds. These would cause an excess of $\GeV$ photons in the spectrum of $\TeV$ blazars. As sufficiently strong EGMFs can dilute the effects of the cascade in the line-of-sight, non-observation of this excess in could be used to place lower limits on the EGMF strength (e.g., \citealt{Oikonomou17}, and references therein). A rigorous analysis of Fermi LAT data for known $\TeV$ blazars has led to a conservative limit of $B \gtrsim 10^{-10}\,\nG$ \citep{2015ApJ...814...20F}; the authors also discuss how additional processes like beam-plasma instabilities \citep{Broderick2012} could suppress the cascade development and render these limits inapplicable. A direct discovery of pair cascades could be possible by observing their spatial and time-like widening of the (potentially transient) blazar emission, known as \textit{pair-halos} and \textit{pair-echoes}, respectively. 
If the EGMF strength turns out to be ${\lesssim 10^{-6}\,\nG}$, future Cherenkov telescopes like CTA, AGIS, or HAWC
would be able to detect such pair-halos or echoes from $\TeV$ blazars and thus to infer further properties of the large-scale EGMFs (e.g., \cite{Neronov09}). EGMFs with a strength significantly above $10^{-6}\,\nG$ cannot be constrained by this method.

For stronger EGMFs, such as are expected in the denser regions of the cosmic web in particular, Faraday RMs of extragalactic sources provide a powerful tool. Recently, statistical
algorithms based on Bayesian inference have been developed to
separate the Galactic and extragalactic contribution \citep{2012A&A...542A..93O}
and to further disentangle the latter into the terms associated with
galaxy clusters, filaments, voids, and sheets
(e.g., \citealt{Vacca2015,2016A&A...591A..13V}). Yet again, in order to detect and constrain the
properties of magnetic fields in the cosmic web, the detailed knowledge of
the Galactic contribution gained by the \imagine\ project will be essential.

\section{Approaches and methods}
\label{s:AM}

\done{We have summarised the scientific background and goals of the \imagine\ project. But to connect the many disparate strands of theoretical and observational knowledge needed requires a statistically powerful and computationally flexible framework. Here, we describe the approach we take to the \imagine\ project. More details as well as a first demonstration are provided in \citet{steininger2018}.}

\editorial{Please note the various subsections and their meaning: 6.1 General discussion of Bayesian methods, NOT specific to GMF inference. 6.2/6.3 Parametric models / non-parametric reconstruction of the GMF - put the GMF specific part here. Same for 6.4, GMF tracers, so observations we use, and 6.5/6.6 Galactic / extragalactic priors applied to our GMF reconstruction.}

\subsection{Introduction to Bayesian methods}

\imagineSW\ is designed to provide a flexible and modular analytic platform based on Bayes theorem that allows the explicit mathematical characterisation of: the constraining data and their uncertainties; the models and their parameters; the posterior likelihood of a given model; any prior information about the model not included in that likelihood; and the quantitative evidence for a given model compared to another.

\subsubsection{Bayesian inference}
\label{sec:Bayes:inference}

The most important thing to understand about Bayesian methods is that they do not introduce a different way of doing statistics, but rather a method of inference in a multi-valued logic appropriate for scientific problems. This goes back to a proof delivered by Cox \citet{Cox1946}, that the desiderata of a logic of plausibilities are fulfilled by the calculus of probability theory, so one can use the terms `plausibility' and `probability' in the same sense. 
Thus Bayes' theorem, commonly written as
\begin{align}
\label{eq:bayes_law}
P\left({\cal M}|{\cal D},{\cal I}\right) &= \frac{P\left({\cal D}|{\cal M}, {\cal I}\right)}{P\left({\cal D}|{\cal I}\right)}
\cdot P\left({\cal M}|{\cal I}\right),
\end{align}
becomes the rule of inference on the \textit{plausibility}
$P \in [0,1]$, assigned to physical entity of interest
${\cal M}$, given a (new) set of empirical data
${\cal D}$ and some background information
${\cal I}$. The meaning of all these quantities in various contexts will be explained in the remainder of this section.

\subsubsection{Parametric vs.\ non-parametric methods
\label{sec:Bayes:para-non-para}}

\done{In \textit{parametric Bayesian methods}, $\mathcal{M}$ is understood as a physical or phenomenological model. Such model prescriptions can be based on purely heuristic assumptions or be the result of a fundamental theory, as described in \ref{ss:Para_phys}.}
\done{Their value lies in physical abstraction, which may be useful to transfer knowledge between related problems, but they are usually oversimplified when it comes to the exact description of a complex structure like the Galactic magnetic field. }

\done{In contrast to this, \textit{non-parametric Bayesian methods} aim to reconstruct  reality to the highest possible accuracy. Here, $\mathcal{M}$ takes the role of a physical \done{field}, e.g., a direct representation}
\done{of the Galactic magnetic field}. 
The main problem here is to recover this field from
the data, which is subject to limitations of measurement and noise.
\done{A field has virtually an infinite number of degrees of freedom, whereas the data provide only a finite number of constraints. Thus the reconstruction problem is heavily under-constrained and ill-defined. This can only be cured by the inclusion of prior information into the inference, which relates the many degrees of freedom.}
\done{To this end, \textit{information field theory} (IFT, \citealt{2009PhRvD..80j5005E,2013AIPC.1553..184E}), the information theory for fields, will be used. Information field theory is a probabilistic formulation of the inverse problem to estimate an unknown field that exploits the rich
variety of mathematical methods developed for quantum field theory.
}

\done{Parametric models describe a subset of the possible field configurations, and therefore parametric models can implicitly be regarded as having strong priors. Nevertheless, their lower dimensionalities can have huge computational and conceptual advantages. For these reasons, the \imagineC\ will investigate both parametric and non-parametric GMF models. In fact, the existing \imagineSW\ pipeline recently presented in \citet{steininger2018} combines a parametric representation of the mean GMF with additional non-parametric realisations of magnetic fluctuations in the inference. The mean field parameters as well as the parameters controlling the magnetic fluctuations can be inferred simultaneously.}

\subsubsection{Likelihood}
\label{sec:Bayes:likelihood}
\done{The likelihood, \done{$P\left({\cal D}|{\cal M, I}\right)$}, is the most crucial element in Bayesian inference.}
\done{It is the probability of obtaining the (observed) data ${\cal D}$ out of all possible data configurations given a specific model realisation $\cal M$ and background information $\cal I$. Knowing the uncertainty of the available data is essential for calculating a likelihood, which acts as a distance measure between measurement and model prediction. The easiest way is to consider the errors communicated by the experiment as part of the data, which strictly speaking they are not as they are already a result of an inference performed by the experimentalists. Non-parametric Bayesian approaches provide sophisticated methods to estimate uncertainties directly from the measurements (see \ref{sss:field_priors}).}

\subsubsection{Prior, posterior and knowledge update}
\label{sec:Bayes:prior}
\done{The leftmost and rightmost terms in \ref{eq:bayes_law}, $P\left(\cal{M}|\cal{D},\cal{I}\right)$ and $P\left(\cal{M}|\cal{I}\right)$, are called the \textit{posterior} and the \textit{prior}, respectively. Both are closely related, as they represent plausibility values assigned to the same entity, $\cal M$. }
\done{The prior is just the assignment before the data is obtained or taken into account, and the posterior is the assignment afterwards. The change from prior to posterior reflects the information the data provided through its likelihood, which is simply multiplied by the prior in Bayes' theorem.}

\done{In practice, there exist several approaches to construct prior distributions. For example, priors may be constructed empirically via a Bayesian knowledge update from previously obtained experimental results. Another possibility is to construct the least-informative priors via the maximum entropy methodology.}

\done{Particularly \jpr{\sout{in}} in non-parametric approaches, informative priors are necessary. The simplest non-parametric priors discourage strong gradients or curvature in the solutions to ensure continuity of the physical fields. More sophisticated priors incorporate the notion of field correlation functions, which either can be calculated from theory (as in CMB science) or have to be determined simultaneously from the data. Thus, in the ideal case, the features introduced to construct sensible non-parametric priors \done{reveal} highly interesting physical quantities. The natural by-products of non-parametric inference are therefore the interpretable scientific results that provide insight and understanding into, e.g., Galactic magnetogenesis.}

\subsubsection{Evidence and background information}
\label{sec:Bayes:evidence}
\done{The last term in Bayes' theorem, given in \ref{eq:bayes_law}, is the denominator on the right side, $P\left({\cal D}|{\cal I}\right)$, which is called the \textit{evidence}. }
\done{Its most important role is to renormalise the joint probability of data and model
\[
P\left(\mathcal{D}, \mathcal{M}|\mathcal{I}\right) = P\left(\mathcal{D}| \mathcal{M},\mathcal{I}\right)\cdot P\left(\mathcal{M}|\mathcal{I}\right)
\]
that arose from multiplying the prior with the likelihood, such that
\[
\int P\left(\mathcal{M}|\mathcal{D}, \mathcal{I}\right)\cdot \dd\mathcal{M}=1.
\]
}

\done{In practice, the evidence allows us to penalise model complexity and choose the simplest model parametrisation able to explain the data. It therefore naturally implements the common understanding of Occam's razor 
in Bayesian model comparisons \citep{2003prth.book.....J}.}

\done{Formally, the evidence, $P\left({\cal D}|{\cal I}\right)$, is the \done{plausibility} of the data ${\cal D}$ in the light of the \done{background} information $\mathcal{I}$, \done{marginalised over all} model parameters.} \done{It might be seen as a measure of the quality of the data, \ie\ their ability to constrain the fundamental approach (or meta-model) contained in ${\cal I}$. Another way to see it is as the likelihood of ${\cal I}$ in the light of the data ${\cal D}$. In this sense,}
\done{it represents information about the plausibility of  $\mathcal{I}$ and permits comparisons between
different formulations of this background information.}

\subsubsection{Numerical approaches to Bayesian inference}
\label{sec:Bayes:num_bayes}

\done{A numerical approach to full Bayesian inference always aims at mapping the entire posterior distribution. This often requires performing numerical searches and integration in very high dimensional parameter spaces, ranging from a few to several thousand dimensions, depending on the parametrisation of models.

In such situations, the Markov Chain Monte Carlo (MCMC) technique is known to be the most efficient numerical technique for approximating high dimensional posterior distributions (see e.g., \citealt{brooks2011handbook}).
The MCMC approach is a class of algorithms generating representative samples of probability distributions (see e.g., \citealt{gelmanbda04}). This is achieved by constructing a so-called Markov chain by sequentially transitioning from one point in parameter space to another. If transitions depend only on the current point and if they obey the requirement of detailed balance, then the equilibrium state of such a chain consists of ergodic samples of the desired target probability distribution. More explicitly, the MCMC algorithm performs a sequential local exploration of parameter space and provides a numerical approximation to the desired target posterior distribution in terms of a multi-dimensional point cloud given as:
\begin{align}
\label{eq:MCMC_approx}
P\left({\cal M}|{\cal D},{\cal I}\right) &\approx  \frac{1}{N} \sum_{i=0}^{N-1} \delta^D\left({\cal M}-{\cal M}_i\right) ,
\end{align}
where $\delta^D(x)$ indicates the Dirac-delta distribution, $M_i$ are sequential model realisations generated by the MCMC process and $N$ is the total number of generated posterior realisations.
It can be shown that for $N \to \infty$, the right hand side of \ref{eq:MCMC_approx} converges to the desired target posterior distribution (see e.g., \citealt{gelmanbda04,brooks2011handbook}). Given such a set of posterior samples in post-processing, one may then perform any desired Bayesian activity such as determining credibility intervals, marginalising over a nuisance parameter or determining statistical summaries such as mean, mode or variance. In particular, one can now easily approximate any posterior-weighted integral over any desired function, $f(\cal M)$, of the model and its parameters by an unbiased estimator given as
\begin{align*}
E\left(f(\cal M)\right) &\approx \frac{1}{N} \sum_{i=0}^{N-1} f({\cal M}_i) .
\end{align*}
A particularly important aspect of the MCMC approach is the fact that the errors of any derived estimators are independent of dimension and scale simply as $1/\sqrt{N}$. These features render the MCMC approach the perfect method to perform Bayesian parameter inference.
}

\subsection{Parametric models \done{of the GMF} \label{sec:GMFpar}}
\label{ss:Para}

\subsubsection{Current heuristic models of the GMF}
\label{ss:Para_heuristic}

There are many parametric models for the large-scale GMF in the literature (e.g., \citealt{sun08,jaffe10,pshirkov11,vaneck11,jansson12b,terral16} ) motivated by observations of the Milky Way and of external galaxies. Most are a combination of a toroidal field (\ie\ no vertical component) and a poloidal field (\ie\ no azimuthal component) to reproduce the observed features described in \ref{ss:OBSGMF}. Most models also parametrise the amplitude as a form of exponential disk, in some cases with spiral arm structures \citep{jaffe10,pshirkov11,vaneck11,jansson12b} or annular regions \citep{sun08} where both the field strength and  direction can vary independently. Further divisions into thin and thick disks, halo, inner and outer Galaxy, a molecular ring, etc.\ attempt to model different regions of the ISM.

The small-scale turbulent component of the field is usually also parametrised statistically, e.g., as a single-scale or 3D Gaussian random field with a parametrised power spectrum. The amplitude of this component usually also follows some sort of exponential disk profile, possibly with spiral arms etc., as in the case of the regular field, but not necessarily the identical morphology. This can then roughly double the number of parameters required (e.g., \citealt{jansson12b}).

The complexity of the parametrisations of such models can clearly vary depending on how many assumptions are made, with several dozen parameters the most that are usually feasible to explore (e.g., \cite{jansson12b, ungerfarrar2017}). As always, the risk of using large numbers of parameters is overfitting and drawing conclusions about the large-scale structure of the field based on a fit that may be perturbed by a small-scale local feature. In principle, the statistics of how the expected small-scale fluctuations affect the fits can be taken into account in the Bayesian framework. In practice, this can be difficult as discussed in \cite{pipXLII}.

Many of the methods in the \imagine\ framework have been applied in previous analyses of the GMF.
\done{Ruiz-Granados et al.}~\cite{ruizgranados:2010} explored several complementary parametrised models with different morphologies and a Bayesian likelihood exploration.
\done{Jansson \& Farrar}~\cite{jansson12b} chose a single parametrised model but one with many physically motivated components and a large number of degrees of freedom. Both of these analyses used the high-resolution data to estimate the variations due to the turbulent field component and to include it in the likelihood.
\done{Jaffe et al.}~\cite{jaffe10} instead explicitly modelled the random field components with an isotropic Gaussian random field and a stochastic anisotropic component.
\done{In}~\cite{pipXLII} those two approaches of handling the random component are compared.
\done{Unger and Farrar}~\cite{ungerfarrar2017} explore an ensemble of physically-motivated updates to the Jansson \& Farrar \cite{jansson12b} model (based in part on Ferri\`ere and Terral \cite{FerriereTerral:2014}). They conclude that the data does currently not discriminate among these model variations, which therefore indicates the degree of uncertainty in our knowledge of the GMF.

We will make use of all of the lessons learned in these analyses by including the ``Galactic variance'' as an observable; comparing with high-resolution simulations of random fields, both Gaussian and non-Gaussian; comparing the ability of different parametrised models to constrain the large-scale features of the GMF; and using the Bayesian information framework to robustly characterise the knowledge gained from the analysis given the assumed parametric forms and priors.

\subsubsection{Physical parametric models}
\label{ss:Para_phys}

\done{However, these models are all heuristic and their topology is based on
observed shapes instead of physics. \done{Ferri\`ere and Terral} \cite{FerriereTerral:2014} were
the first to develop analytical models for both disk and halo components of GMFs and \done{in} \cite{terral16} \done{these were applied} to the
Milky Way. Although these models still include only regular field components and are \done{clearly} oversimplified, they do present a significant step forward, as the field topology is based on physics, in contrast to the published heuristic models for which field topology consists of geometrical components based on observations.}

The next advance is building parametric GMF models based on dynamo
physics.
This theory
is aimed mainly at the explanation of the origin of GMFs and their typical properties. To obtain a parametrised model, a kinematic
(linear) solution \done{of \ref{MFDeq} is derived} for $\vec{V}$, $\alpha$ and $\eta_\mathrm{t}$,
independent of $\vec{B}$. The linear nature of the solution is not
restrictive in the present context since its aim is just to provide a
convenient functional basis for any magnetic configuration. On the other
hand, Chamandy~et~al.~\cite{CSSS14} show that a wide class of non-linear
solutions is well approximated by the marginally stable eigenfunction (\ie\ that is obtained for $\partial\vec{B}/\partial t=0$).

\begin{figure}
 \centering
 \includegraphics[width=0.9\columnwidth,height=0.42\textheight]{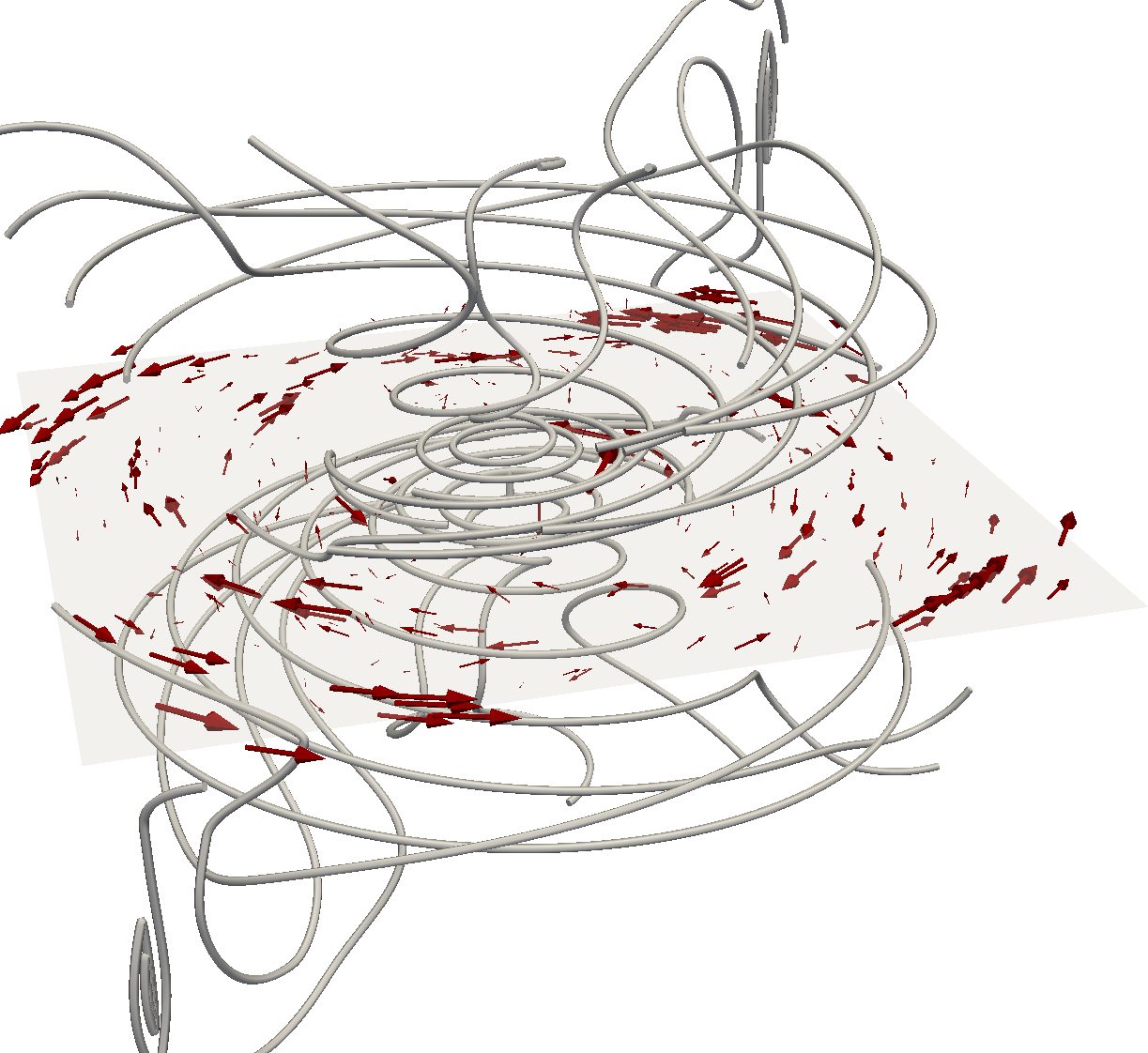}
 \caption{A GMF model based on the expansion,  in 
 free-decay magnetic modes, of axially symmetric solutions of kinematic 
 mean-field dynamo equations for the Galactic disc and halo. For  
 illustration, we have selected a solution that has a symmetric 
 (quadrupolar) halo field combined  with a quadrupolar disc field 
 that has two  reversals at arbitrarily chosen 
 Galacto-centric distances $7$ and $12\,\kpc$. The domain
 shown is $17\,\kpc$ in size. Magnetic field lines (shown in grey) 
 are seeded uniformly along a major diagonal through the box. Arrows show 
 the magnetic field at positions randomly sampled within a $2.5\,\kpc$-thick 
 slice around the Galactic mid-plane (indicated by the semi-transparent 
 surface); their length is proportional to the magnitude of the magnetic field.}
 \label{GMF_3D}
\end{figure}

In a thin disc with radius $r$ and height $h$, 
a series of successive approximations in aspect ratio $\epsilon = h/r$
can be used to represent \ref{MFDeq} as:  a system of local partial differential equations in $z$ at a fixed Galacto-centric radius $r$, whose coefficients are functions of $r$; and a scalar equation for the magnetic field strength at a given $r$. Solutions to each approximation are obtained in an explicit form as an infinite series of orthogonal functions. The series can be truncated in order to achieve the desired level of detail in the result. In particular, such solutions can include:
deviations from axial symmetry due to spiral arms and other factors; radial and vertical large-scale velocity
components; inhomogeneity and anisotropy of the turbulent transport coefficients; magnetic field reversals; etc.
For the quasi-spherical Galactic halo, a similar solution to \ref{MFDeq} is obtained in spherical coordinates
in the form of expansion over the free-decay modes
obtained for $\vec{v}=0$ and $\alpha=0$. Analytical
expressions are available for these modes,
expressed in spherical harmonics, so that the solution is again represented by an explicit form with a configurable
level of detail.
An example of a global (large-scale) magnetic configuration 
in the Galactic disc and halo produced with this model is shown in \ref{GMF_3D}.

In order to provide specific models of the GMF, this solution should be extended to include a
variety of complex physical effects,
such as the multi-phase structure of the ISM,
spiral arms, the effects of cosmic rays on the gas
dynamics, etc.
Some of these effects have already
been included, e.g., the effects of galactic
outflows \citep{BvRDBS01},
and there is some progress towards a better
  understanding of the role of the multi-phase ISM structure in
  galactic dynamos \citep{EGSFB16}, but much more needs to be done.

Another difficulty, of a more fundamental nature, is that any theory of the random magnetic
fields can only be statistical in nature, predicting
their ensemble-averaged properties, whereas
the GMF is just one realisation of the ensemble.
However, the existing models of the GMF,
especially those based on the mean-field dynamo theory,
can provide a physically motivated field prior for a non-parametric Bayesian
analysis of observations. 

\subsubsection{Challenges for parametric modelling}

We have discussed the variety of models for the large-scale GMF that have been published, each having been optimised to match a subset of the available data (\ref{ss:Para_heuristic}). Though there are other suitable observables, the most commonly used tracers of the large-scale fields are: the Faraday RMs of Galactic pulsars and extragalactic radio sources;  diffuse synchrotron emission in total and polarised intensity from radio to microwave frequencies; and diffuse dust polarisation in the microwave and sub-millimetre bands.  As discussed in detail in \citet{pipXLII}, the problem
of determining the magnetic structure of the Milky Way in sufficient detail
remains under-determined due to
degeneracies in the parameter space. The ill-constrained distribution of thermal electrons and the paucity of pulsars with reliable distance measures make it difficult to study the 3D structure of the fields with RMs.  Confusion from additional emission components in the microwave bands such as free-free or anomalous microwave emission (aka ``spinning dust'') make it impossible to use the synchrotron total intensity at high frequencies.  Faraday depolarisation effects cut off the visible polarisation in the lower frequency radio bands at a polarisation ``horizon'' of a few $\kpc$ \citep{uyaniker:2003}.  Unknown variations in the synchrotron spectral energy distribution make it difficult to combine radio and microwave observations.  
Models of dust polarisation must account for its correlation with synchrotron polarisation \citep{choi15,PIRXXII2015}, as well as the correlation of the GMF with the filamentary structure of matter \citep{PIRXXXII2016}. 
The structure of the local magnetic field on scales of a few
$100\,\pc$, e.g., the Local Bubble \citep{lallementetal2014}
is currently not taken into account in existing GMF models. However, 
the analysis of the \planck\ dust polarisation maps stress the need 
to consider this local contribution in order to model polarisation observations away from the Galactic plane \citep{pipXLII,PIRXLIV2016,alves2018}.

Widely used models such as that of Jansson \& Farrar \citet{jansson12b} are often taken at face value without an understanding of how these systematic uncertainties affect the deceptively good fit obtained for each parameter.  More data on the current observables, such as diffuse synchrotron polarisation at intermediate frequencies and better pulsar sampling, as well as new observables such as CR deflection information will help resolve some of these degeneracies, but it will remain necessary to build into any model fitting an understanding of the remaining uncertainties.  The Bayesian framework of \imagineSW\ will allow us to do this explicitly and thereby to gain a better understanding of which parameters can be meaningfully constrained with the available data.

\subsection{Non-parametric reconstruction of the GMF}
\label{ss:non-para-models}

Non-parametric 3D GMF models do not yet exist, and are a challenge due to the large number of degrees of freedom involved compared to the number of constraining data points. It is a second goal of the \imagineC\ to enable non-parametric reconstructions
\done{of the GMF}. All data and instrument descriptions collected by the \imagineC\ will also be directly applicable to this more challenging goal.

A non-parametric reconstruction of the GMF requires that basically every point in 3D space carries its own magnetic field vector. As this is computationally infeasible to handle, only a pixelised version of the 3D GMF configuration can be stored. However, even such a finite resolution can easily mean that a billion field values not only have to be handled but also need to be determined from a much smaller set of observables. This is a so-called ill-posed inverse problem, which requires additional constraints or regularisation best provided by prior information such as:  the GMF's solenoidality;  the typical scaling relations of magnetic fluctuations in MHD turbulence; and the fact that the GMF is the result of an operating galactic dynamo that obeys the MHD equations and is driven by the kinetic energy of the ISM.
The Bayesian framework of \imagineSW\ will allow us to test the ability of these priors to constrain non-parametric models to provide information that will be complementary to the parametric studies.

 \subsection{Observational input}
\label{ss:OI}

In \ref{sss:oot}, we described the physics of the observational tracers of Galactic magnetic fields. Here, we specifically list the available data sets for each tracer that we will use in \imagineSW.

\subsubsection{Rotation measure and Faraday depth}

The first and most straightforward data set to compare the GMF models against is that of RMs of extragalactic point sources. For this, we use a map of Galactic Faraday rotation reconstructed based on catalogues of extragalactic point source RMs \citep{2012A&A...542A..93O}. Pulsar RMs have additional value due to their measurable distances within the Galaxy, however, there are relatively few (${\sim}\,200$) pulsars with accurately known distances and RMs \citep{yaoetal2017}. In principle, the Faraday depth of diffuse Galactic synchrotron emission (e.g., \citealt{Iacobelli:2013b, jelicetal2014, vanecketal2017}) gives magnetic field information as well. However, due to local, small-scale structure and observational biases, this tracer may not be very useful in practice. RM and Faraday depth data depend on a reliable thermal electron density model (e.g., \citealt{cordeslazio2002, gaensleretal2008, schnitzeler2012}) and a relativistic electron density model (e.g., \citealt{strongmoskalenko1998, kissmann2014, evolietal2017}).

\subsubsection{Synchrotron emission}

With a relativistic electron density model in place, it is straightforward to calculate the synchrotron emission expected from the Milky Way given a certain GMF model. The modelled synchrotron maps at various frequencies can be directly compared to existing observational maps (see e.g., the compilation by \citealt{deoliveiracostaetal2008}). The complication of local structures needs to be dealt with, e.g., by including them in a model or masking them out.
At very low frequencies, Faraday rotation by small-scale, local magnetic field structures will dominate, but at higher frequencies the global field should be constrainable. The all-sky polarisation surveys from the WMAP \citep{pageetal2007} and \planck\ satellites \citep{planck2015XXV} are effectively free of Faraday rotation. Lower-frequency surveys that can be used that include Faraday rotation are, e.g., the all-sky survey at $1.4\,\GHz$ \citep{wollebenetal2006, testorietal2008}, the Southern sky survey S-PASS at $2.3\,\GHz$ \citep{carrettietal2013}, or the full-sky C-BASS survey at $5\,\GHz$ \citep{kingetal2010}.

\subsubsection{Polarisation of starlight}

The optical and near-infrared polarisation of stars is a potentially powerful constraint for the Galactic magnetic field, as shown in pilot studies by \citet{paveletal2012} and \citet{pavel2014}. The catalogue of Heiles \citet{heiles2000} and the Galactic Plane Infrared Polarisation Survey (GPIPS, \citealt{clemens2012}) combined with stellar distances from Gaia can be used to start with. A huge increase in stellar polarisation data is expected from ongoing optical stellar polarisation surveys: IPS \citep{magalhaesetal2005}, SOUTH-POL \citep{magalhaes2014} and Pasiphae, which will be combined with stellar distances and extinction measurements from Gaia \citep{lindegrenetal2016}. The modelling of stellar polarisation will go together with that of the dust 
distribution and extinction properties in 3D. Such models based on extinction measurements \citep{lallement2018,green2018}, and
on spectroscopy of diffuse interstellar bands \citep{zasowski2015}, are already available.

\subsubsection{Polarised dust emission}
\done{The same elongated dust particles that selectively absorb optical and near-infrared starlight, re-emit the absorbed polarised light in the far-infrared and sub-millimetre regimes. Therefore, the expected polarised emission from the chosen 3D dust model can be calculated and compared to observed all-sky maps of dust emission at various wavelengths from \planck\ \citep{PIRXXI2015}.
\done{These maps provide sensitive, full-sky observations of the total and polarised emission up to $353\,\GHz$.}
}\done{Again, fully exploiting these data requires an accurate 3D model of the dust grain distribution.}
The observed asymmetry between $E$ and $B$ mode polarisation and their correlation with the total dust intensity
\citep{PIRXXX2016,PIRLIV2018} provide specific constraints and challenges to the GMF modelling. 

\subsubsection{Ultra-high energy cosmic ray deflections}
\label{sec:UHECR_tracer}

\afterpage{
\addtocounter{footnote}{-1}
\begin{figure}[tb]
\centering
\includegraphics[width=0.9\columnwidth]{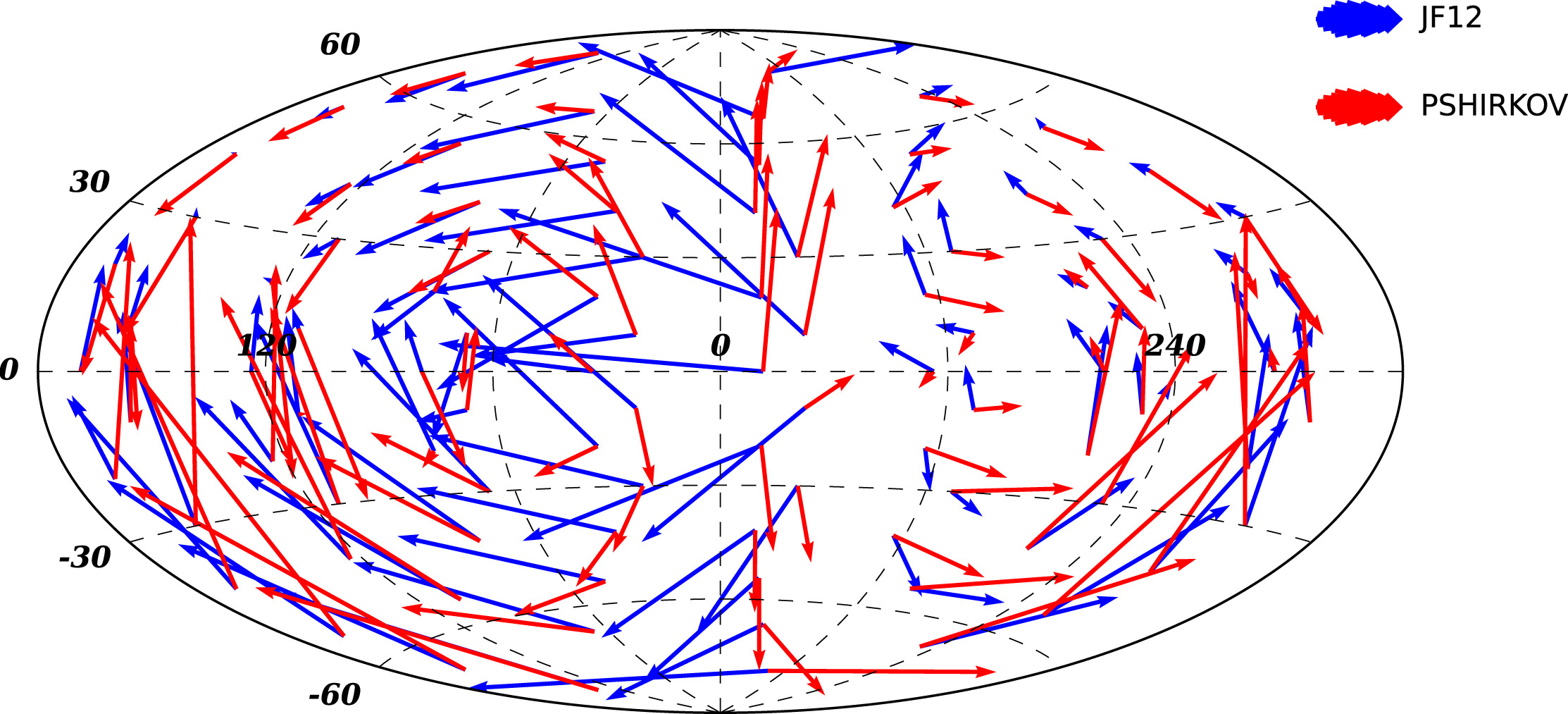}
\caption[Comparison of deflection angles of UHECRs with rigidity $E/eZ = 10\,\EV$ predicted by two published models of the GMF.]{Comparison of deflection angles of UHECRs with rigidity
$E/eZ = 10\,\EV$ predicted by two published models of the GMF: Pshirkov et~al. \citep{pshirkov11} and Jansson \& Farrar (JF12) \citep{jansson12b}. \jpr{\sout{Image credit: S.~Mollarach and E.~Roulet \citep{2018PrPNP..98...85M}.}\footnotemark}
\label{fig:defl_diff}}
\end{figure}
\footnotetext{\jpr{Reprinted from Progress in Particle and Nuclear Physics, Vol. 98, S.~Mollerach and E.~Roulet, \textit{Progress in high-energy cosmic ray physics}, figure~15, p.~107, \textcopyright\ (2018), with permission from Elsevier.}}
}

UHECRs can act as test particles to probe the 3D GMF structure in a
unique way: unlike many other tracers, they probe the orientation of
the transverse field component, and their total deflection is not
simply a line-of-sight integral. This uniqueness becomes clear when
looking at the inconsistency in the predictions for systematic UHECR
deflections by different parametrisations for the GMF, which are both
optimised at standard GMF tracers (\ref{fig:defl_diff}, see also
\citet{Farrar:2012gm}). Including UHECRs in the GMF likelihood would require connecting various UHECR anisotropy predictions outside the Galaxy with the observed arrival direction distribution at Earth, and promising techniques to do this efficiently have been developed \citep{Winchen:2013}. Considering information on the particle rigidity for each air shower would increase the discriminatory power of the likelihood, and experimental developments give hope to have detailed information on $X_{\rm max}$ soon available for a large number of air showers \citep{2017PhRvD..96l2003A}.

UHECR deflections in magnetic fields inside and outside of the Galaxy can be calculated by numerical codes like \crpropathree\ \citep{Batista:2016yrx}, which can take any given GMF and EGMF structure as a 3D grid. To probe the GMF structure, UHECRs with moderate deflection angles ${\sim} 10\degree$ will be most useful, provided that the original anisotropy is significant on intermediate scales. First hints in this direction have been delivered from anisotropy analysis at the highest measured energies \citep{Abbasi:2014lda,AugerStarburst}, which suggests that indeed UHECRs with $E/eZ > 10\,\EV$ exist. Together with theoretical UHECR priors and constraints from multi-messenger signals (see \ref{UHECR:prior}), it is therefore the variety of the information contained in the UHECR spectrum that can help to break degeneracies in GMF model optimisations.

\subsection{Galactic priors} 
\label{ss:galactic_priors}

As discussed in \ref{sec:Bayes:prior}, a prior reflects our knowledge before the observation of the data and is particularly important for non-parametric GMF modelling (see \ref{ss:non-para-models}). Here we describe how we will formulate such priors for our galaxy.

\subsubsection{Parameter priors \label{sec:GMF_par_priors}}

Parametrised models for the magnetic field structure already encode the constraints applied to describe the problem, e.g., by assuming a spiral disk morphology, \done{or requiring $\nabla\cdot\vec{B} = 0$}.

In heuristic GMF models, priors on the values of individual parameters can be either non-informative or constrained;  the latter can make the analysis computationally feasible or reduce degeneracies in parameter space.  They may also be used to include empirical constraints obtained from earlier optimisations, potentially also with different parametrisations. Because the results depend on the choices made at the start regarding both analytic model formulations and parameter priors, it will be important to test a number of different possibilities and to explore their mutual consistency.

A somewhat different situation occurs in dynamo models of the GMF, which are primarily based on an underlying physical process rather than a set of morphological assumptions. As described in \ref{ss:Para_phys}, these are based on
the expansion of the large-scale magnetic field over a basis of
orthogonal eigenfunctions of the mean-field dynamo equation.
Since the functional basis is complete,
any magnetic field, whether or not produced by the dynamo, can be
represented as a superposition of the eigenfunctions. Therefore,
an alternative use of these functions  is to represent
any magnetic configuration of interest in terms of a tractably small
number of parameters. The resulting magnetic field will be physically
realisable, as it is by construction a solution of the induction equation. This then permits us to construct a prior for it. Different modes have different growth or decay rates. It is much more likely to observe a strong excitation of a growing mode than of a decaying mode for the simple reason that the latter are eliminated by the dynamics. Thus the eigenvalues of the dynamo modes can be turned into probabilistic statements of GMF field configurations.

\subsubsection{Field priors}
\label{sss:field_priors}

For a non-parametric GMF reconstruction, priors are essential to tame
the (in principle) unbound number of degrees of freedom beyond
the finite number of constraints provided by the data. Here, certain
field smoothness assumptions can be applied, where the field
roughness can be learned from the data themselves, as was shown to
work well theoretically
\citep{2010PhRvE..82e1112E, 2011PhRvD..83j5014E,2011PhRvE..84d1118O,2013PhRvE..87c2136O,2016arXiv161208406E} and in practice
\citep{2011A&A...530A..89O, 2012A&A...542A..93O, 2015A&A...581A..59J, 2015A&A...575A.118O, 2015A&A...581A.126S, 2016A&A...590A..59G, 2016arXiv160504317G,2016A&A...586A..76J,2016A&A...591A..13V}.  Furthermore, magnetic fields are solenoidal,
$\vec{\nabla} \cdot \vec{B} = 0$, which removes a third of the
field's degrees of freedom in Fourier space. Finally, the GMF is a
result of an MHD dynamo, obeying the MHD equations and being driven
by ISM gas flows.
One can thus use the dynamo equation to quantify for each possible field configuration how transient it is. As it is much more likely to observe at a given instant a long lasting configuration than a very transient one, the former should be assigned a larger prior probability.

Considering short duration configurations is still necessary given the current theoretical understanding. The unknown correlation structure, or equivalently, the unknown field power spectrum, then represents a latent, or hidden, variable that should be inferred simultaneously but that also guides the field reconstruction. For example, the GMF power spectrum is shaped by the dynamo equation and the mean velocity field. The properties of these can also be regarded as latent variables that could and should be inferred as well.
Power spectra estimation and the exploitation of such spectra to improve
reconstructions are by now standard operations of IFT algorithms, 
as used for constructing all-sky maps from gamma-ray data (D$^3$PO; \citealt{2015A&A...574A..74S,2015A&A...581A.126S}), for radio
interferometry (RESOLVE; \citealt{2015A&A...581A..59J, 2016A&A...586A..76J,
2016arXiv160504317G}), and for non-parametric
Galactic tomography \citep{2016A&A...590A..59G}.

\subsection{Extragalactic priors}
\label{ss:extragalactic_priors}

\done{Information gathered from outside our galaxy can also be important to formulate Bayesian priors to be used in \imagine, as outlined in the following.}

\subsubsection{Galaxies} 
\label{sss:Galaxies}

Idealised cosmological simulations of
forming galaxies will never {\it exactly} reproduce the phases and amplitudes of
the CR and magnetic field distributions in our galaxy. However, they will be invaluable
tools that enable predicting the (higher-order) statistics of these
distributions, which can then be used to provide useful Bayesian priors for
inference estimates such as the \imagine\ framework presented here.
An important aspect of a useful Galaxy model is the mass distribution, both near the Galactic centre and within the DM halo.
In particular, such questions as the plausible or admissible form of the rotation curve at small Galacto-centric distances depend on the
inner DM density profile and the presence of a cusp. The structure of the DM halo and its substructures (e.g., the abundance and properties of sub-halos) can affect the large-scale properties of the interstellar gas flows and distribution. Another aspect of galaxy evolution that affects the formulation of the priors is related to magnetic field estimates in Mg\textsc{ii} absorbers at high redshift \citep{Bernet2008}.
In order to translate them into useful constraints on the Galactic magnetic field (in particular, in the Galactic halo), a clear understanding of galactic evolution is required.

Another rich source of information is observations of nearby galaxies. Detailed observations of synchrotron emission and Faraday rotation are available for dozens of spiral galaxies. The spatial distributions of these tracers along and across galactic discs and in their haloes (\citealt{2015A&ARv..24....4B} and references therein)  provide rich constraints on the likely and unlikely forms of the Galactic magnetic field.

\subsubsection{UHECR source and arrival direction distribution}
\label{UHECR:prior}

Although the origin of UHECRs is still an unsolved problem, the enormous energies of these particles constrain the options to a few physically reasonable scenarios. All of them make unique predictions for the relation of the spectrum, chemical composition, and arrival directions, which predictions can be used as priors for UHECR deflection studies. The key lies in the so-called Hillas plot,
which summarises the possible sources of UHECRs in a diagram of size $R$ vs.\ magnetic field strength $B$ through the relation
$BR \ge E_{\rm max}/eZ$ \citep{1984ARA&A..22..425H}.

The traditional way to interpret this plot is to see it as an empirical collection of distinct source classes. Listed from smallest to largest linear sizes, these are mainly:  pulsar wind shocks \citep{Kotera:2015pya,Lemoine:2015ala}, tidal disruption events \citep{AlvesBatista:2017shr,Biehl:2017hnb,Guepin:2017abw}, gamma-ray bursts \citep{1995PhRvL..75..386W,1998AIPC..428..776R,Globus:2014fka,2017arXiv171209984Z}, radio galaxies and AGN \citep{1993A&A...272..161R,1993A&A...273..377R,2008arXiv0808.0349R,Farrar:2008ex,2018JCAP...02..036E}, and large-scale accretion shocks \citep{1995ApJ...454...60N,Kang:1995xw,Kang:1996rp}. All these models make predictions for the spectrum, composition and anisotropy of UHECRs depending on parameters constrained by other astronomical observations and can be compared in our analysis via their Bayesian evidence in view of all available constraints.

Another way is to see the Hillas plot as the result of a general, underlying non-thermal process responsible for the production of UHECRs and to relate the contribution of various source types to the properties of cosmological structure formation. This approach allows us to understand some general features of UHECRs, in particular their observed maximum energies as well as the structure of the Hillas plot \citep{Rachen:2015Texas}, but also allows us to construct a parametrised prior for UHECR origin if combined with reconstructions of DM-driven large-scale structure formation.

Observational constraints on extragalactic UHECR priors may be constructed from multi-messenger information like the arrival directions of cosmic neutrinos \citep{Aartsen:2014gkd} or gamma-rays \citep{Acero:2015hja}.
It should be noted, however, that the relation between UHECR emission and the production of secondaries is highly non-trivial and requires detailed modelling of source properties like acceleration sites, mechanisms and target densities, as well as orientation effects. \done{Independent of such ambiguities regarding direct gamma-ray emission from the sources, future $\TeV$ gamma-ray observatories may be able to trace UHECR production in the Universe via the cosmogenic gamma-ray halos that arise from interactions of these particles with cosmic backgrounds \citep{GA05,KAL11}}

\section{IMAGINE software design}
\label{s:SD}

We have discussed extensively the scientific background and goals of the \imagine\ project and the mathematical and scientific tools we will use. Here, we given an overview of the software components of the \imagineSW\ pipeline we have developed. For further details, \done{tests, and} its first application, see \citet{velden2017thesis,steininger2018}.

\subsection{Design overview} 


\begin{figure}[tbp] \centering
		\begin{tikzpicture}[>=stealth,thick,every node/.style={font=\relsize{0.9}, draw, shape=rectangle,rounded corners,align=center, anchor=center}, scale=0.966]

		\node at (0, -1.5) (samp) {Sampler};
		\node at (-4, -1.5) (prior) {Prior};

		\node at (4, -1.5) (rep) {Repository};
		
		\node at (0, 0) (pipe) {Pipeline};
    	\node at (4, 0) (sample) {Sample};

		\node at (-4, 2) (bgen) {Galaxy-Generator};
		\node at (-4, 4) (bfield) {Galaxy-Instance};

		\node at (0, 4) (obsgen) {Observable-\\Generator};
		\node at (4, 4) (obs) {Observables};
		
        \node at (0, 2) (data) {Data};
		\node at (4, 2.5) (like) {Likelihood};

		\draw[->] (pipe) to[out=120,in=-90] (bgen);
		\draw[->] (bgen) to[out=90,in=-90] (bfield);
		\draw[->] (bfield) to[out=0,in=180] (obsgen);
		\draw[->] (obsgen) to[out=0,in=180] (obs);
		\draw[->] (obs) to[out=-90,in=90] (like);
		\draw[->] (data) to[out=0,in=180] (like);
		\draw[->] (like) to[out=-90,in=60] (pipe);
		\draw[<->] (pipe) to[out=0,in=180] (sample);
    	\draw[<->] (sample) to[out=-90,in=90] (rep);
		\draw[->] (prior) to[out=0,in=180] (samp);

		\draw[<->] (samp) to[out=90,in=-90] (pipe);
		
		\end{tikzpicture}

      \caption{The structure of the \imagine\ data processing and interpretation.}
      \label{fig:imagine_structure}

\end{figure}

The structure of the \imagineSW\ pipeline is shown in \ref{fig:imagine_structure}. The observables (\ref{ss:OI}) are compared to model predictions (\ref{ss:Para} and \ref{ss:non-para-models}) and the likelihood (\ref{sec:Bayes:likelihood}) assessed including prior information (\ref{ss:galactic_priors} and \ref{ss:extragalactic_priors}) using Bayesian statistics (\ref{sec:Bayes:inference}).

\imagineSW\ is a pipeline framework rather than a software application, since one of its core concepts is extensive modularity. In the context of \imagineSW, the Galaxy is described by a set of physical fields, e.g., the Galactic magnetic field, the thermal electron density, or the dust density.
Those fields each have independent degrees of freedom with which they are fully described.
For a certain parametric GMF model, for example, a few dozen parameters might be sufficient.
In contrast, a non-parametric GMF model has as many degrees of freedom as voxels, e.g., $2^{24}$ for a box with a resolution of $256 \times 256 \times 256$.
From the pipeline's point-of-view, it is irrelevant which constituents make up this abstract Galaxy or how many dimensions make up the parameter space of the constituents.

The individual parts of the \imagineSW\ framework, shown in \ref{fig:imagine_structure}, are described in the following.
The central object for the parameter space exploration is the \texttt{pipeline} object that
coordinates the likelihood calculations and evaluations.
Its most important counterpart is the \texttt{sampler}, which the \texttt{pipeline} provides with an abstract \texttt{likelihood} and a \texttt{prior} functional.
Currently, \pymultinest\ \citep{2014A&A...564A.125B}, based on the nested sampling algorithm of \citet{skilling2006}, is used as the default sampler.
The \texttt{sampler} triggers the likelihood evaluation for each point in a normalised parameter space through the following steps.
\begin{itemize}
\item First, the parameter set given by the \texttt{sampler} is used to generate a certain Galaxy model realisation.
In practice this means creating the dynamic constituents our abstract Galaxy model consisting of, e.g., a Galactic magnetic field, a thermal electron density field, a dust density field, etc.;
\item The \texttt{Galaxy-instance} is then processed by an \texttt{observable-generator} that simulates physical \texttt{observables} like rotation measures or maps of the thermal dust emission;
\item That model is then assessed by a \texttt{likelihood} functional that compares the simulated \texttt{ob\-servables} to measured \texttt{data};
The \texttt{pipeline} can store the resulting likelihood value in a \texttt{repository} to allow for caching and post-processing of the calculated data before it is forwarded to the \texttt{sampler}.
\end{itemize}

Although a \texttt{Galaxy-instance} is basically determined by the parameters of its constituents, the constituents can possess random components.
For example, the Galactic magnetic field has a large stochastic component.
Because of this, instead of just creating one \texttt{Galaxy-instance}, the \texttt{Galaxy-generator} actually creates a whole ensemble of instances that share the same parametrisation.
After being processed by the \texttt{observable-generator}, the ensemble is then used by the \texttt{likelihood} functional to estimate the Galactic covariance.

\begin{figure}[tbp] \centering
        \begin{tikzpicture}[>=stealth,thick,every node/.style={font=\relsize{0.9}, draw, shape=rectangle,rounded corners,align=center, anchor=center},scale=0.966]

        \node at (3, 2) (imagine) {\LARGE IMAGINE};

        \node at (0, 0) (nifty) {NIFTy};
        \node at (3, 0) (hammu) {Hammurabi};
        \node at (6.5, 0) (pymultinest) {PyMultiNest};

		\node at (7, -2) (multinest) {MultiNest};
        \node at (5, -3) (healpix) {HEALPix};
        \node at (-2.5, -2) (numpy) {NumPy};
        \node at (2.5, -2.5) (fftw) {FFTW3};
        \node at (0, -3) (d2o) {D2O};

        \draw[->] (healpix) to[out=90,in=-60] (nifty);
        \draw[->] (numpy) to[out=90,in=-120] (nifty);
        \draw[->] (fftw) to[out=90,in=-80] (nifty);
        \draw[->] (d2o) to[out=90,in=-100] (nifty);

        \draw[->] (fftw) to[out=90,in=-100] (hammu);
        \draw[->] (healpix) to[out=90,in=-80] (hammu);

        \draw[->] (multinest) to[out=90,in=-85] (pymultinest);

        \draw[->] (nifty) to[out=90,in=-120] (imagine);
        \draw[->] (hammu) to[out=90,in=-90] (imagine);
        \draw[->] (pymultinest) to[out=90,in=-60] (imagine);

        \end{tikzpicture}

      \caption{The building blocks of the \imagineSW\ pipeline framework.}
      \label{fig:imagine_building_blocks}

\end{figure}
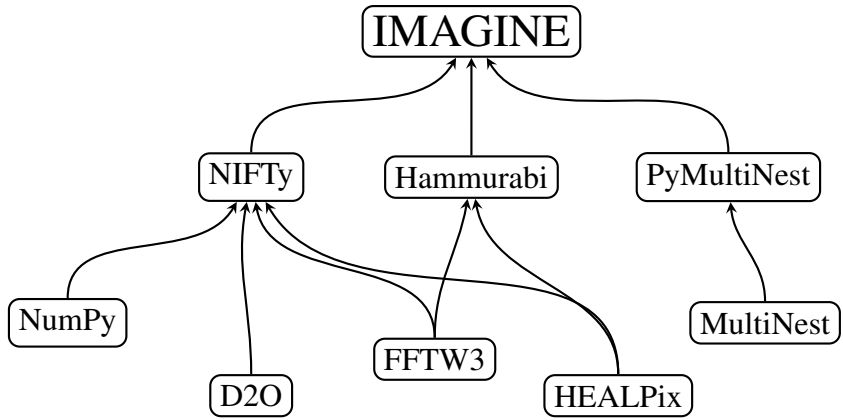

The computational effort to process a whole ensemble of dozens or even hundreds of \texttt{Galaxy\-instance}s for each position in parameter space can be mitigated via parallelisation.
For convenient data processing in general and efficient data parallelisation in particular, the \imagineSW\ framework is built on the software packages \niftythree\ \citep{2017arXiv170801073S} and \dtwoo\ \citep{SGBE16}, respectively.
An overview of the framework's building blocks is given in \ref{fig:imagine_building_blocks}.

\subsection{\done{The hammurabi code}} 

The \hammurabi\ code\footnote{\url{http://sourceforge.net/projects/hammurabicode/}} \citep{waelkens:2009} is an astrophysical simulator based on 3D models of the components of the magnetised ISM such as magnetic fields, thermal electrons, relativistic electrons, and dust grains.  It performs an efficient line-of-sight integration through the simulated Galaxy model using a \healpix\footnote{\url{http://healpix.sourceforge.net}}-based nested grid to produce observables such as Faraday rotation measure and diffuse synchrotron and thermal dust emission in full Stokes $I$, $Q$ and $U$, while taking into account beam and depth depolarisation as well as Faraday effects.  This modular code allows new analytic models for each component to be added easily, or alternatively an external file can be given that specifies the model in a binary grid.  The public version already includes relatively simple field models such as the axisymmetric spiral with reversed ring of Sun et~al. \citep{sun08} and more complicated models such as Jansson \& Farrar \citep{jansson12b}, which includes spiral arm segments and an X-shaped halo field.  For small-scale turbulence, the code includes a Gaussian random field simulator to add a simple random component to any analytic mean-field model.  This code is the basis of one of the engines of the \imagineSW\ machinery, converting parametric models for the magnetised ISM into observables that can then be compared to data in the likelihood evaluation.

An updated version, \hammurabiX\ (Wang et al., in preparation), is currently under development in order to achieve the higher computing performance required by \imagineSW.
Previously in \hammurabi, the generation of the anisotropic component of the random field as well as the rescaling of the field strength following various parametric forms lead to unphysical divergence.
Now we propose two novel solutions for simulating random/turbulent magnetic field.
On Galactic scales, a triple Fourier transform scheme is proposed to restore the divergence-free condition via a cleaning method.
Alternatively, in the local Solar neighbourhood,
a vector-field decomposition scheme is capable of simulating a more detailed turbulent field power-spectrum.
In addition to these new field generators, the simulation accuracy has been improved with a calibrated trilinear interpolation algorithm and the implementation of Simpson's $1/3$ rule in line-of-sight integration.
The input and output control and \python\ wrapper are further developed to ensure an efficient interface with \imagineSW.

Furthermore, in the future, \hammurabiX\ will be extended for application to IFT tomographic GMF reconstructions, \ie\ non-parametric modelling as described in \ref{ss:non-para-models}. For this purpose, the derivative of the simulated observables with respect to the field values has to be computed, \ie\ the linearised response of the simulated data to small model changes. For the reverse inference from the data to the field configuration, we also require the adjoint matrix to express how a mismatch between the real and simulated data sets tends to pull the model. Since parametric models can be regarded as being embedded within the space of all non-parametric models, an efficient gradient-based optimisation of parametric models will also become feasible using this extension; the linearised response operator then uses the differential relation between the 3D field configuration and the model parameters.

\subsection{Sampler} 

An important part of the \imagine\ project is drawing robust and accurate conclusions from observations subject to a variety of systematic as well as stochastic uncertainties. Depending on the quality of the real data and the complexity of the respective data models, we expect to encounter counter-intuitive interdependencies and degeneracies among inference quantities, nuisance parameters and noise properties. Quantifying these effects requires the joint and fully self-consistent treatment of all these quantities within a rigorous information theoretical inference approach.

Various MCMC techniques (discussed generically in \ref{sec:Bayes:num_bayes}) are known to efficiently explore high-dimensional parameter spaces (see e.g., \citealt{brooks2011handbook}). The numerical and statistical efficiency of this algorithm crucially depends on the design of proper transition kernels. Optimal transitions, obeying the detailed balance criterion, can be designed with the Metropolis-Hastings procedure \citep{txt:Metropolis,hastings1970}, the basis of almost any modern MCMC algorithm. Different MCMC approaches mostly only differ in the design of transition kernels that are optimal for different target posterior distributions. The \texttt{sampler} module of the \imagineSW\ pipeline can easily exploit any state-of-the-art MCMC technique by simply connecting it with the corresponding software library. In particular the \imagineSW\ pipeline can exploit available inference packages such as \pymc\ \citep{pymc2015}, \pymcthree\ \citep{pymc3_2016} and \stan\ \citep{stan:2017}. This immediately enables a user of the \imagineSW\ pipeline to access various sampling algorithms ranging from random walk Metropolis-Hastings and Gibbs sampling to Hamiltonian Monte Carlo sampling \citep{txt:Metropolis,hastings1970,Geman_1984,1987PhLB..195..216D,Neal2012}.

As discussed above, the literature provides a plenitude of rival models that need to be compared and judged with respect to the data. In Bayesian parlance, we will have the data decide among models. This task is different from parameter inference, since it requires a judgement of the validity of models independently of their respective model parameters. This task therefore amounts to performing Bayesian model comparison.

A technical challenge of Bayesian model comparison is the numerical determination of the evidence. This is an active field of research and has not yet been conclusively solved. Nevertheless, for moderate dimensionality the task of numerically estimating the evidence of the posterior distribution can be solved by performing nested sampling (see e.g., \citealt{Skilling04}). The \imagineSW\ pipeline can readily use several publicly available software packages implementing the nested sampling algorithm \citep{2014A&A...564A.125B,2015MNRAS.450L..61H,2015MNRAS.453.4384H}. In particular the present implementation of the \imagineSW\ framework uses the \pymultinest\ library for the evidence calculation \citep{steininger2018}. From a users' perspective, Bayesian model comparison can now easily be performed by running the \imagineSW\ pipeline with several different models of the magnetic field and comparing estimated evidence values for respective models, as discussed in \ref{s:AM}.

\subsection{NIFTy}
\label{ss:nifty}
\done{Information field theory \citep{2009PhRvD..80j5005E} is information theory for fields and therefore the ideal language to phrase and solve non-parametric field inference problems
(see e.g., \citealt{2010PhRvE..82e1112E, 2011PhRvD..83j5014E,2011PhRvE..84d1118O, 2013PhRvE..87c2136O, 2016arXiv161208406E}) that performs well in practice (e.g., \citealt{2011A&A...530A..89O, 2012A&A...542A..93O, 2015A&A...581A..59J,2015A&A...575A.118O, 2015A&A...581A.126S, 2016A&A...590A..59G, 2016arXiv160504317G,2016A&A...586A..76J,2016A&A...591A..13V}).
}

These numerous applications of IFT to real world problems were possible thanks to the \textit{numerical information field theory} (\nifty, \citealt{2013A&A...554A..26S, 2013ascl.soft02013S}) package. \nifty\ permits the direct implementation of IFT equations by providing the concepts of spaces, fields living over the spaces, and operators acting on the fields in a transparent, abstract and object oriented manner to the programmer, thereby alleviating the need to think about the properties of a chosen space pixelisation.
In order to be prepared for the task of reconstructing the 3D GMF, \nifty\ was recently parallelised \citep{SGBE16} and rewritten \citep{2017arXiv170801073S} such that it \jpr{\sout{can}} can comfortably deal with vector valued fields.
 For a non-parametric GMF reconstruction, the handling of large datasets needs to be mastered as well as the vector nature of the magnetic fields. For both challenges, preparations are under way.

\section{IMAGINE Consortium}
\label{s:IC}
The \imagineC\ was conceived as an informal collaboration of astronomers, astrophysicists, experimental and theoretical physicists and applied mathematicians who share
interest and expertise in the broad range of problems related to
Galactic magnetic fields.

The collaboration that has developed into the \imagineC\ started with the 
International Team Meeting, \textit{``Bayesian modelling of the Galactic Magnetic Field
Constrained by Space and Ground-Based Radio-Millimetre and Ultra-High Energy Cosmic Ray Data''} hosted in 2014/15 by the International Space Science Institute in Bern, Switzerland\footnote{\url{http://www.issibern.ch/teams/bayesianmodel}}. The next milestone was a
workshop hosted in 2017 by the Lorentz Center in Leiden, the Netherlands\footnote{\url{http://www.lorentzcenter.nl/lc/web/2017/880//info.php3?wsid=880&venue=Snellius}}, titled \textit{``A Bayesian
View on the Galactic Magnetic Field: From Faraday Rotation to Ultra-High Energy Cosmic Ray Deflections''}, where the \imagineC\ was formally founded. The goals of the consortium are:

\begin{itemize}

\item \done{to reveal the 3D structure of the Galactic magnetic field;}

\item \done{to achieve a comprehensive understanding of the non-thermal ISM joining the observational and theoretical efforts
of the participants in the studies of the ISM, GMF and cosmic rays;}

\item \done{to support theoretical and experimental efforts to identify the sources of UHECRs;} 

\item \done{to develop novel methods to infer the GMF and other ISM components that exploit and fuse information from observations, theory, and simulations;}

\item \done{to develop, maintain and promote the \imagineSW\ pipeline as a standard Bayesian framework;}

\item to encourage interaction between distinct research areas and cross-disciplinary knowledge transfer. 

\end{itemize}
\done{The \imagineC\ is led by a PI team consisting of Fran\c{c}ois Boulanger, Torsten En{\ss}lin, Marijke Haverkorn, J\"org H\"orandel, Tess Jaffe, Jens Jasche, J\"org Paul Rachen and Anvar Shukurov, and} is open to new members. 
Information about the consortium activities and information on how to become a member will soon be available at the \imagine\ webpage.\footnote{\url{https://www.astro.ru.nl/imagine/}}

\acknowledgments
We thank the International Space Science Institute in Bern, Switzerland, and the Lorentz Center in Leiden, the Netherlands, for the hospitality and financial support \done{that has led to the founding of the \imagineC.} We thank Sebastian Hutschenreuter for providing \ref{fig:PMF}\jpr{, and Franco Vazza for providing \ref{fig:egmf_Hackstein} based on simulations done with the ENZO code\footnote{\url{http://cosmosimfrazza.myfreesites.net/erc-magcow}} \citep{0264-9381-34-23-234001}.} TE and TS acknowledge partly support by the DFG Cluster of Excellence ``Origin and Structure of the Universe'' and by the DFG Research Unit 1254 ``Magnetisation of Interstellar and Intergalactic Media -- The Prospects of Low-Frequency Radio Observations''. AF, LFSR and AS thank STFC (ST/N00900/1) and The Leverhulme Trust (RPG-2014-427) for funding. PG and CP acknowledge support by the European Research Council under ERC-CoG grant CRAGSMAN-646955. 
BR-G acknowledges \jpr{support from the European Union's Horizon 2020 project} RADI\jpr{O}FOREGROUNDS \jpr{\sout{H2020's project}} under grant  agreement number 687312. GS is supported by the DFG through collaborative research centre SFB 676 ``Particles, Strings and the Early Universe'' and by the Bundesministerium
für Bildung und Forschung (BMBF) through grant 05 A17GU1. TS was supported by the Studienstiftung des deutschen Volkes. AvV acknowledges financial support from the NWO astroparticle physics grant WARP.

\bibliographystyle{JHEP_IMAGINE}
\bibliography{imagine}
\end{document}